\documentclass[english,aps,pra,amsmath,twocolumn,preprintnumber,superscriptaddress,showpacs,notitlepage,nofootinbib]{revtex4-1}
\usepackage[T1]{fontenc}
\usepackage[latin9]{inputenc}
\makeatletter
\usepackage{mathptmx,newtxtext,newtxmath,xspace}
\usepackage{amsbsy,bm,bbold}
\usepackage{graphicx,color,xcolor,epsfig,rotate}
\usepackage{fancyhdr}
\usepackage{gensymb}
\usepackage[colorlinks=true,
            linkcolor=blue,
            urlcolor=blue,
            citecolor=blue]{hyperref}
\usepackage{soul}
\usepackage{braket}
\setstcolor{red}
\newcommand{\iu}{{i\mkern1mu}}

\makeatother

\usepackage{babel}
\begin{document}

\title{Meron, skyrmion, and vortex crystals in centrosymmetric tetragonal magnets}
\author{Zhentao~Wang}
\affiliation{Department of Physics and Astronomy, The University of Tennessee,
Knoxville, Tennessee 37996, USA}
\affiliation{School of Physics and Astronomy, University of Minnesota, Minneapolis, Minnesota 55455, USA}
\author{Ying~Su}
\affiliation{Theoretical Division, T-4 and CNLS, Los Alamos National Laboratory,
Los Alamos, New Mexico 87545, USA}
\author{Shi-Zeng~Lin}
\affiliation{Theoretical Division, T-4 and CNLS, Los Alamos National Laboratory,
Los Alamos, New Mexico 87545, USA}
\author{Cristian~D.~Batista}
\affiliation{Department of Physics and Astronomy, The University of Tennessee,
Knoxville, Tennessee 37996, USA}
\affiliation{Neutron Scattering Division and Shull-Wollan Center,Oak Ridge National Laboratory, Oak Ridge, Tennessee 37831, USA}
\date{\today}

\begin{abstract}
The recent experimental confirmation of a transformation between meron and skyrmion topological spin textures in the chiral magnet Co$_8$Zn$_9$Mn$_3$~[S.-Z. Lin {\it et al.}, Phys. Rev. B {\bf 91}, 224407 (2015); X.~Z.~Yu {\it et al.}, Nature {\bf 564}, 95 (2018)] confirms that the skyrmion crystals discovered in 2009~[S.~M\"{u}hlbauer {\it et al.}, Science {\bf 323}, 915 (2009)] are just the tip of the iceberg. Crystals of topological textures, including skyrmions, merons, vortices, and monopoles, can be stabilized by combining simple physical ingredients, such as lattice symmetry, frustration, and spin anisotropy. The current challenge is to find the combinations of these ingredients that produce specific topological spin textures. Here we report a simple mechanism for the stabilization of meron, skyrmion, and vortex crystals in centrosymmetric tetragonal magnets. In particular, the meron/skyrmion crystals can form even in  absence of magnetic field. The application of magnetic field leads to a rich variety of topological spin textures that survive in the long wavelength limit of the theory. When conduction electrons are coupled to the spins, these topological spin textures twist the electronic wave functions to induce Chern insulators and Weyl semimetals for specific band filling fractions.
\end{abstract}
\pacs{~}

\maketitle

\section{Introduction}
Inspired by the work of Herman Helmholtz~\cite{Helmholtz67}, William Thomson proposed
in 1867 that  atoms could be vortices in ether~\cite{Thomson67}. While later experiments put this proposal out of business, thinking of topological solitons as emerging building blocks or artificial atoms is  very appealing. Indeed, more recent developments that started around the 1960's~\cite{Skyrme61,Skyrme88} have demonstrated that nature has plenty of room for finding updated versions of  ether~\cite{Wilczek2011}. The ``ether of quantum magnets'' is the vector field of magnetic moments, whose topological solitons can be regarded as emergent mesoscale atoms~\cite{Bogdanov89}. Like real atoms, these solitons form crystal structures  dictated by symmetry, anisotropy, and competing  interactions.

Periodic arrays of topological spin textures typically arise from the superposition of small-${\bm Q}$  spirals propagating along symmetry-related directions (multi-${\bm Q}$ ordering), whose wavelength is dictated by the competition between ferromagnetic (FM) and antiferromagnetic interactions. In chiral magnets, the FM interaction competes against  the antisymmetric  component of the effective exchange tensor, known as Dzyaloshinskii-Moriya interaction~\cite{Dzyaloshinsky1958,Moriya60b}. However, the selection of a small-${\bm Q}$ spiral ordering is not enough to stabilize topological spin textures because the superposition of multiple spirals requires additional energetic considerations. For instance,  it is known that
threefold symmetric lattices favor the formation of skyrmion crystals (SkXs) (triple-${\bm Q}$ ordering) induced by a magnetic field parallel to the $c$ axis, because the Ginzburg-Landau (GL) free energy can include  terms of the form $({\bm S}_{\bm 0} \cdot {\bm S}_{{\bm Q}_1})  ({\bm S}_{{\bm Q}_2} \cdot {\bm S}_{{\bm Q}_3})$  (${\bm Q}_1+{\bm Q}_2+{\bm Q}_3=0$ because the three ordering wave vectors differ by $2\pi/3$ rotations). This simple consideration explains why the vast majority of magnetic SkXs have been found by applying magnetic field along a {\it threefold} symmetry axis of different materials~\cite{Muhlbauer2009}.

While less common,  square skyrmion and meron crystals have been recently reported in chiral~\cite{Lin2015,Karube2016,Yu2018_meron}, polar~\cite{Kurumaji2017,kurumaji2019direct}, and  centrosymmetric tetragonal magnets~\cite{Khanh2020_squareSkX}. The observation of SkXs in fourfold symmetric lattices forces us to think about alternative stabilization mechanisms.
Here, we report a guiding principle for the formation of meron cystals (MXs), which are also SkXs, in fourfold (tetragonal) lattices \emph{even in the absence of magnetic field}. The merons form a square lattice and the magnetic unit cell includes four merons with a net skymion charge  $n_\text{sk}=\pm 1$ (MX-I) or $n_\text{sk}=\pm 2$ (MX-II). The phase diagram also includes a field-induced vortex crystal (VtX).
Remarkably, this rich phase diagram is obtained from a very simple model for centrosymmetric magnets that only includes competing easy-axis and  compass anisotropies.

The different phases of the phase diagram are obtained by minimizing the energy over all the possible spin configurations for a fixed magnetic unit cell, whose size is dictated by the ordering wave vector.
We also provide approximated expansions of the relevant spin configurations that include the fundamental
Fourier components ${\bm q}={\bm Q}_{\nu}$ and a few higher harmonics. Our analysis is complemented with the derivation of the GL theory that describes the long wavelength limit of the microscopic model, allowing us to demonstrate the universal character of the phase diagram.
We also  analyze the
anomalous Hall response of itinerant electrons coupled to the local magnetic moments~\cite{Zener51}, and the possible realization of a magnetic Weyl semimetal~\cite{Funk1260} induced by the MX-II phase.

{The rest of paper is organized as follows: In Sec.~\ref{sec:model}, we introduce the microscopic model and compute the corresponding $T=0$ phase diagrams. In Sec.~\ref{sec:GL}, we {introduce a Ginzburg-Landau theory for the continuum limit of the model}, and discuss the stabilization condition of the MX-II phase. In Sec.~\ref{sec:Hall_2D}, we analyze the topological Hall effect that results from coupling the MX-I and MX-II spin textures to   conduction electrons. In Sec.~\ref{sec:Weyl}, we discuss the generation of Weyl points in vertically stacked  layers of MX-II spin textures. In Sec.~\ref{sec:summary}, we summarize the  {main conclusions of the paper}. Appendix~\ref{sec:methods} {provides details of the variational methods that were employed to obtain the $T=0$} phase diagrams. Appendix~\ref{sec:appendix_Q} includes the analysis of the ordering wave vectors
{in the isotropic limit of the model}.  {A further analysis of the skyrmion charge in the meron crystal phases is provided in Appendix~\ref{sec:chargeQ}}. Appendix~\ref{sec:fourier} includes the Fourier analysis of the states that were obtained from the unbiased (fixed unit cell) variational method. Appendix~\ref{sec:stability_MXII} includes  details of the stability analysis of the MX-II phase. Finally, we present the symmetry analysis of the electronic bands in Appendix~\ref{sec:symmetry}.}

\begin{figure*}[tbp]
\centering
\includegraphics[width=1\textwidth]{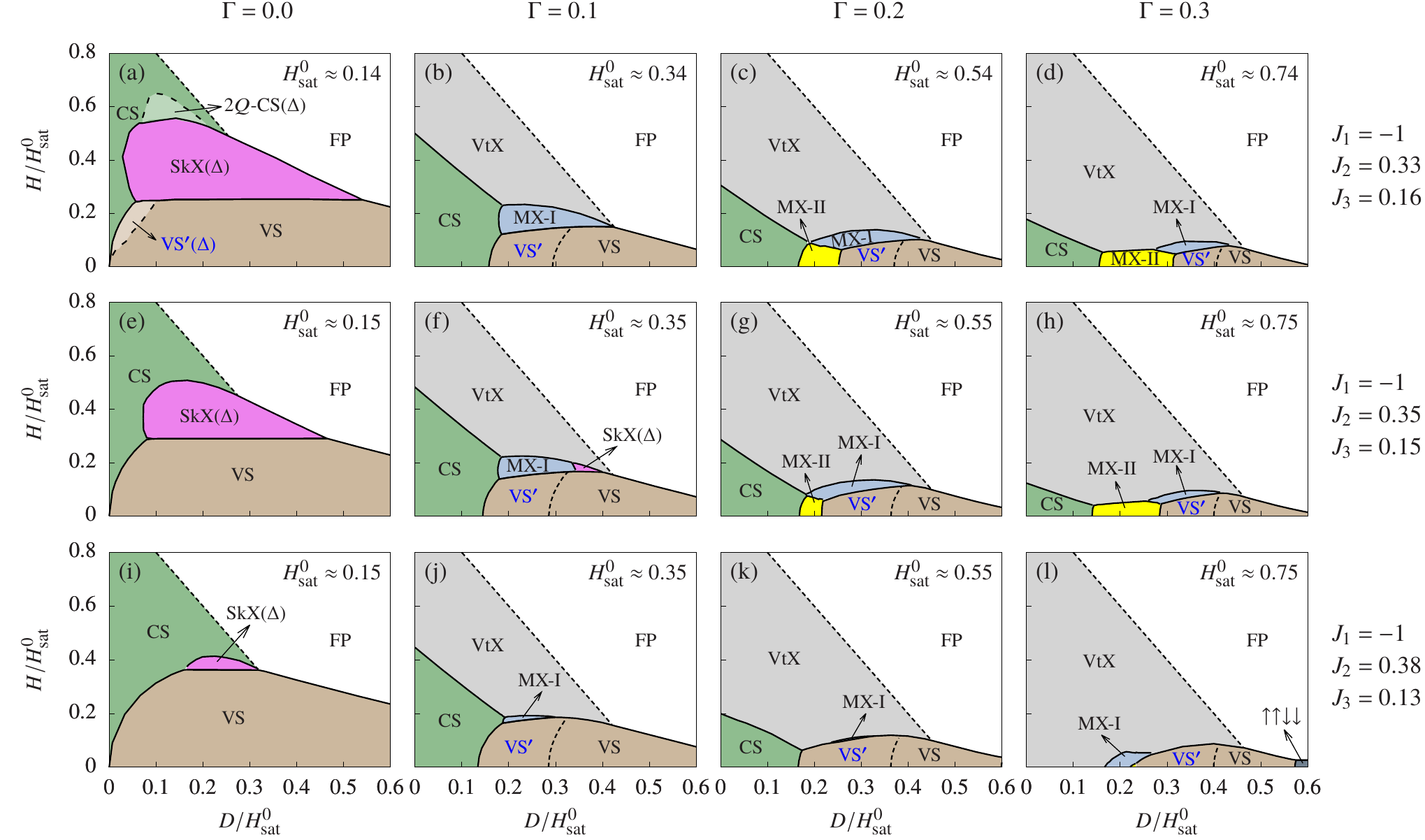}
\caption{Phase diagrams of the microscopic model Eq.~\eqref{eq:model}
at $T=0$. The values of $\{J_{1},J_{2},J_{3}\}$ are denoted on
the right-hand side of each row (see also Fig.~\ref{fig:phd_wavevector}), and the values of
$\Gamma$ are denoted on top of each column. {The solid (dashed)  lines  indicate first- (second-) order phase transitions.}}
\label{fig:phd}
\end{figure*}

\section{Microscopic Model}\label{sec:model}
We consider the {\it classical} Heisenberg model on the square lattice:
\begin{equation}
\begin{split}\mathcal{H} & =\sum_{\langle ij\rangle} J_{ij} \bm{S}_{i}\cdot\bm{S}_{j} -\Gamma\sum_{i}\left(S_{i}^{x}S_{i+\hat{x}}^{x}+S_{i}^{y}S_{i+\hat{y}}^{y}\right) \\
 & \quad -D\sum_{i}\left(S_{i}^{z}\right)^{2}-H\sum_{i}S_{i}^{z},
\end{split}
\label{eq:model}
\end{equation}
where $J_{ij}=$\{$J_{1}$, $J_{2}$, $J_{3}$\} are the Heisenberg interactions
up to the third-nearest neighbor, $\Gamma$ is the compass anisotropy, $D$ is the single-ion anisotropy, and $H$ is the magnetic field along the $z$ axis. In this paper, we consider the case of  FM nearest-neighbor Heisenberg exchange  ($J_1<0$).
The magnitude of the spins is fixed by the normalization condition $|\bm{S}_{i}|=1$. The compass
anisotropy can either be generated by the spin-orbit coupling (SOC) or by the dipolar interaction.

\begin{figure*}[tbp]
\centering
\includegraphics[width=1\textwidth]{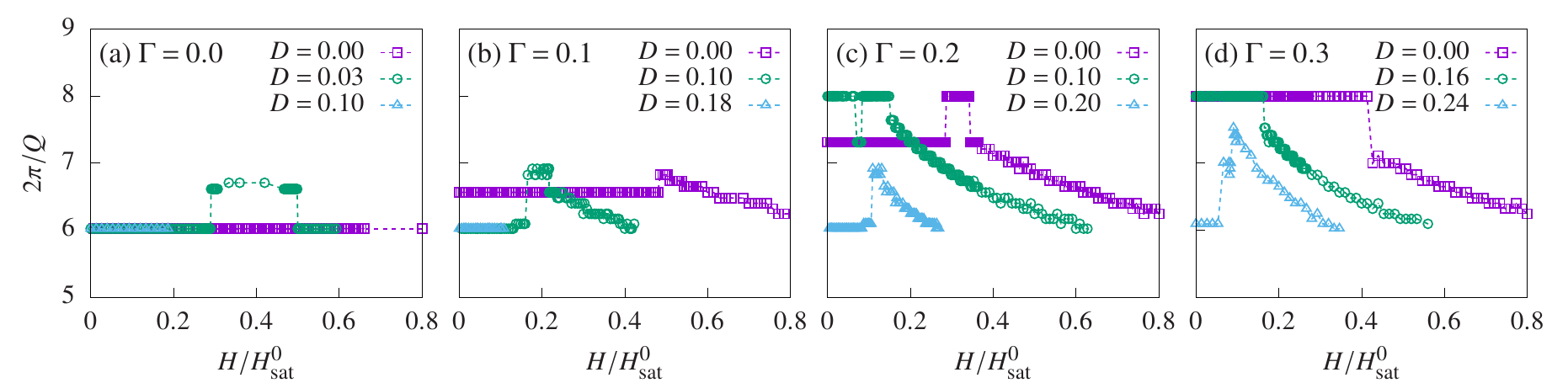}
\caption{{Wavelength of the spin textures obtained from the microscopic model Eq.~\eqref{eq:model}
at $T=0$ (see Appendix~\ref{sec:methods} for details). Here we set $J_1=-1$, $J_2=0.35$, $J_3=0.15$, and the values of $H_\text{sat}^0$ are denoted on the corresponding panels in Fig.~\ref{fig:phd}. Note that some of the points have a small error bar, which can be inferred from the noise of the corresponding curves, because of the limited  resolution of the minimization algorithm. }
}
\label{fig:Q}
\end{figure*}

\begin{figure*}[tp]
\centering
\includegraphics[width=1\textwidth]{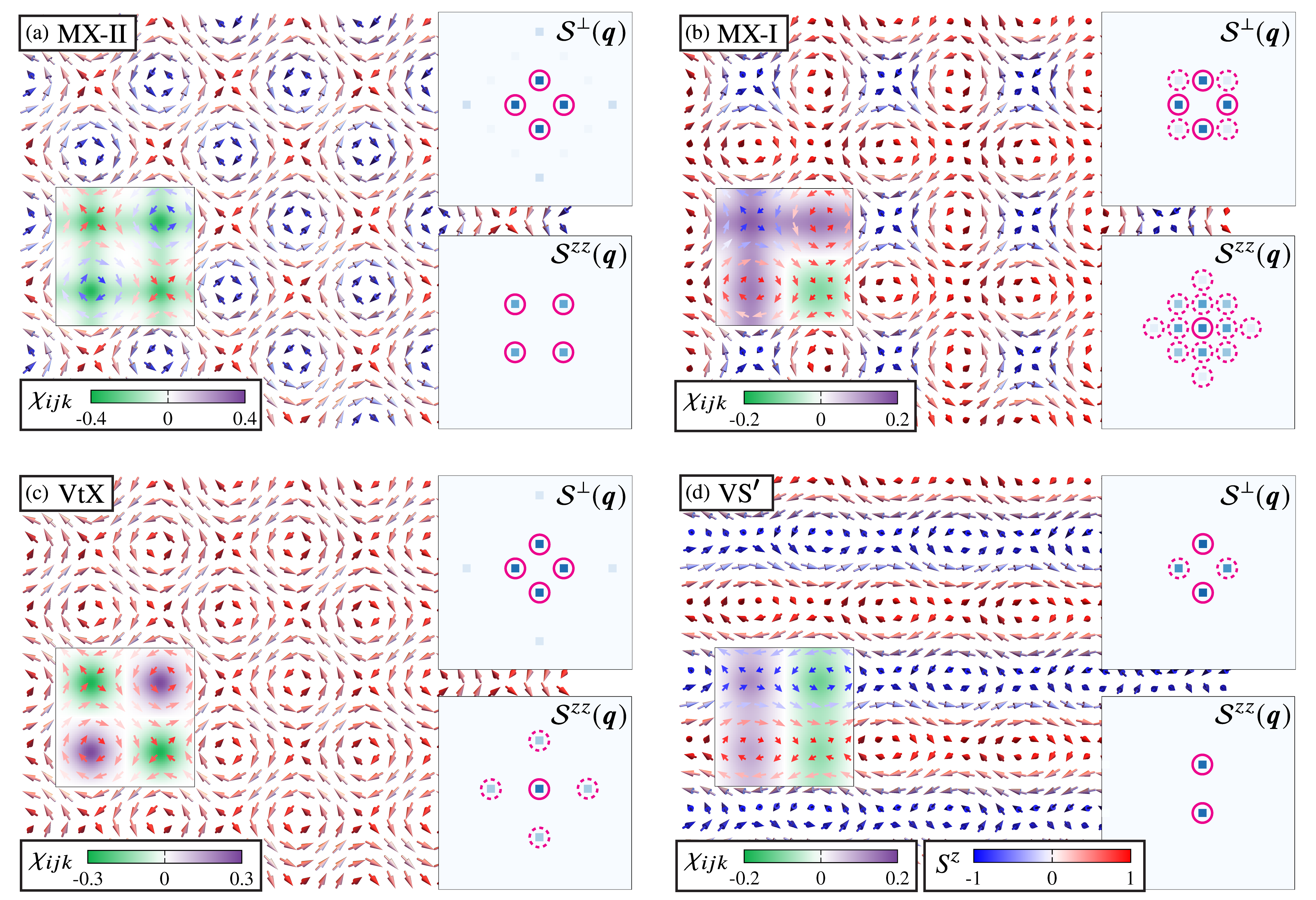}
\caption{Spin configurations of the double-$\bm{Q}$ states. The insets show the in-plane ($\mathcal{S}^{\perp}$)
and out-of-plane ($\mathcal{S}^{zz}$) static spin structure factors
in the first BZ (the intensities are colored in log scale). The solid (dotted) circles highlight the dominant
(subdominant) peaks for $q\le2Q$.
On the left part of each panel, we have also colored the scalar chirality $\chi_{ijk}$ in the magnetic unit cell.}
\label{fig:Spin-plot}
\end{figure*}

The characteristic length scale of the spin
structure is determined by the competition between different symmetric exchange interactions~\cite{Okubo12,Leonov2015,Lin2016_skyrmion,Hayami16,Batista16,Gao2020}.
In the absence of anisotropies and magnetic field, the ordering
wave vector $\bm{Q}$ is obtained by minimizing the exchange interaction
in momentum space:
\begin{equation}
\begin{split}J(\bm{q}) & =J_{1}\left(\cos q_{x}+\cos q_{y}\right)+2J_{2}\cos q_{x}\cos q_{y}\\
 & \quad+J_{3}\left(\cos2q_{x}+\cos2q_{y}\right).
\end{split}
\label{eq:exchange_kspace}
\end{equation}
In the long wavelength limit, we have
\begin{equation}
J(\bm{q}) \simeq I_0 - \frac{I_1}{2} q^2 + \frac{I_2}{2} q^4 - I_3 q^2_x q^2_y,
\end{equation}
with $I_0=2(J_1+J_2+J_3)$ and
\begin{equation}
I_{1}  =J_{1}+2J_{2}+4J_{3}, \;
I_{2}  =\frac{J_{1}+2J_{2}+16J_{3}}{12}, \;
I_{3}  =I_2 - \frac{J_{2}}{2}.
\end{equation}

Depending on the sign of the quartic anisotropy, $I_3$, the competing exchange interactions lead
to spiral phases with ordering wave vectors $\bm{Q}=(\pm Q_0,0)$ or
$\bm{Q}=(0,\pm Q_0)$ with $Q_0=\sqrt{I_1/2I_2}$ for $I_3 <0$, and $\bm{Q}=(\pm Q_0,\pm Q_0)/\sqrt{2}$
with $Q_0=\sqrt{I_1/( 2 I_2-I_3)}$ for $I_3 >0$.\footnote{Such analysis is valid in the long wavelength limit. The corrections for finite $|\bm{Q}|$ can be found in Appendix~\ref{sec:appendix_Q}.}
In {both situations}, the ground state can either be a single-$\bm{Q}$ spiral or a  multi-$\bm{Q}$ structure,
depending on the values of the spin anisotropies and the magnetic field.
Once these extra terms are included in
the Hamiltonian, the optimal value of $\bm{Q}$ can also change. We note, however, that  this change has been found
to be negligible for several cases of interest~\cite{Leonov2015,Wang2020_RKKYskx}.

The different rows of Fig.~\ref{fig:phd} include the $T=0$ phase diagrams of $\mathcal{H}$
for three different sets of exchange interactions, \{$J_{1}$, $J_{2}$,
$J_{3}$\}  that produce relatively small values of $Q$
and  quartic anisotropy:
$I_3=0.02 |J_1|$, $I_3=0$ and $I_3 \approx-0.037 |J_1|$. The different columns
correspond to four different values of the compass anisotropy $\Gamma$.
{As shown in Fig.~\ref{fig:Q}, the inclusion of these extra terms indeed modifies the ordering wave number.}

Since $I_3$ and $\Gamma$ are the main sources of tetragonal anisotropy, the first column of
Fig.~\ref{fig:phd} [Figs.~\ref{fig:phd}(a), \ref{fig:phd}(e), and \ref{fig:phd}(i)] describes cases with weak lattice anisotropy.
Thus, it is not surprising that the phase diagrams are similar to the ones
obtained in the continuum limit of the {\it isotropic theory}~\cite{Lin2016_skyrmion}.
The ordering wave vectors \{$\bm{Q}_{1}$, $\bm{Q}_{2}$, $\bm{Q}_{3}$\}
differ by $120^{\circ}$, implying that SkX($\Delta$) has hexagonal symmetry
in spite of the  underlying tetragonal atomic lattice.
The conical spiral (CS) is just a canted cycloidal spiral induced by the effective easy-plane anisotropy produced by the applied magnetic field. In contrast, the easy-axis anistropy term $D>0$ favors a proper-screw or vertical spiral (VS).
The additional two phases,  2$\bm{Q}$-conical
spiral [2$\bm{Q}$-CS($\Delta$)]\footnote{The name 2$\bm{Q}$-CS($\Delta$) follows the convention of Refs.~\cite{Leonov2015,Wang2020_RKKYskx},
where $\mathcal{S}(\bm{q})$ actually has intensity at all \{$\bm{Q}_{1}$,
$\bm{Q}_{2}$, $\bm{Q}_{3}$\}, but the spectral weight at the third
$\bm{Q}$ can be quite small compared to the other two.},
and the vertical spiral with in-plane
modulation {[}VS$^{\bm{\prime}}$($\Delta$){]}, have also been reported
for centrosymmetric magnets with hexagonal symmetry~\cite{Okubo12,Leonov2015,Lin2016_skyrmion,Hayami16,Batista16,Wang2020_RKKYskx}.

The phase diagram is qualitatively different for $\Gamma /|J_1| \geq 0.1$ because
the compass anisotropy term is strong enough to enforce the tetragonal anisotropy on the spin textures.
The propagation wave vectors, $\bm{Q}_{1}=(Q,0)$ and $\bm{Q}_{2}=(0,Q)$ of the conical and the VSs  are now pinned along the principal $x$ and $y$ directions of the tetragonal lattice.\footnote{Note also that a new
$\uparrow\uparrow\downarrow\downarrow$
state appears for large enough anisotropy [see Fig.~\ref{fig:phd}(l)].}
However, the most remarkable effect of the
$\Gamma$ term is the stabilization of double-${\bm Q}$ orderings that are superpositions of two
proper screw spirals with propagation wave vectors $\bm{Q}_{1}$ and $\bm{Q}_{2}$.
In particular, the MX-II,  MX-I, and VtX phases, shown in Figs.~\ref{fig:Spin-plot}(a)--\ref{fig:Spin-plot}(c),
have the same intensity in the spin structure factor at both wave vectors; while the vertical spiral with in-plane modulation
(VS$^{\bm{\prime}}$) has different intensities [Fig.~\ref{fig:Spin-plot}(d)], implying that it is not invariant under a $90^\circ$ rotation.

The magnetic unit cell of the  MX-II state includes four merons [Fig.~\ref{fig:Spin-plot}(a)], whose
skyrmion charge adds up to $|n_{\text{sk}}|= 2$
[Figs.~\ref{fig:Spin-plot}(a) and \ref{fig:contour}(a)], meaning that the MX-II phase is simultaneously a double-SkX state with  net scalar spin chirality $\chi_{ijk}= \bm{S}_i\cdot \left(\bm{S}_j \times \bm{S}_k \right)$.
Note that the scalar chiralities of the merons have the same sign because a change of sign in the vector spin chirality $\hat{z}\cdot \left(\bm{S}_j \times \bm{S}_k \right)$ (vorticity) is always accompanied by a sign change of $S^z_{i}$.
The MX-I phase is induced by magnetic field, but it
can also exist at zero magnetic field [Fig.~\ref{fig:phd}(l)]. Its
magnetic unit cell includes four merons with a total skyrmion
charge $|n_{\text{sk}}|=1$ [Figs.~\ref{fig:Spin-plot}(b) and
\ref{fig:contour}(b)], implying that this phase is also a SkX.
In this case, one of the merons has opposite scalar chirality relative to the other three  because
of a sign change in $S^z_{i}$ near its core.
It is interesting to note that the MX-I state
is topologically equivalent to the spin configurations that have been observed
in  chiral crystals (Fig.~1(e) of Ref.~\cite{Yu2018_meron}) and
centrosymmetric magnets (Fig.~1(d) of Ref.~\cite{Khanh2020_squareSkX}).
The finite skyrmion charges of the  MX-I and MX-II phases make them qualitatively different from the MXs with $n_\text{sk}=0$ that have been reported in previous works~\cite{Yi2009,Chen2016}.

The VtX state can in principle exist all the way from zero magnetic field up to the saturation (Fig.~\ref{fig:phd}).
Similar to the MX-I and MX-II states, this texture includes four vortices in each magnetic unit
cell. The main difference is that their topological charges cancel with each other:
$n_{\text{sk}}=0$ [Figs.~\ref{fig:Spin-plot}(c) and
\ref{fig:contour}(c)].
The origin of this cancellation is easy to understand  at high  fields, where $S^z_{i}$  does not change signs and the net scalar chirality becomes proportional to the net vector chirality.
Hexagonal versions of this VtX phase have also been reported for  frustrated {\it quantum} magnets below the saturation field~\cite{Kamiya2014,Wang2015_vortex}.

The in-plane spin components are almost \textit{identical}
for the MX-I, MX-II, and VtX states [Figs.~\ref{fig:Spin-plot}(a)--\ref{fig:Spin-plot}(c)],
making them equally good candidates for the double-$\bm{Q}$ magnetic textures
revealed by Lorentz transmission electron microscopy~\cite{Yu2018_meron,Khanh2020_squareSkX}.
We must then rely on other measurements to discriminate between the three possibilities.
As shown in Figs.~\ref{fig:Spin-plot}(a)--\ref{fig:Spin-plot}(c),
the $zz$ component of the static spin structure factor, $\mathcal{S}^{zz}(\bm{q})$,
is different for the three  double-$\bm{Q}$ orderings.
In other words, a polarized neutron or x-ray diffraction experiment can
identify the nature the double-$\bm{Q}$ state.
We also note that $\mathcal{S}(\bm{Q}_{1}+\bm{Q}_{2})\approx0$ for
the VtX state [Fig.~\ref{fig:Spin-plot}(c)], meaning that the spectral weight
at the first harmonic, $\bm{Q}_{1}+\bm{Q}_{2}$, is not always a good criterion to distinguish
a multi-domain single-$\bm{Q}$ phase from a double-$\bm{Q}$ state.
Recently, a double-$\bm{Q}$ state has been reported in the centrosymmetric tetragonal magnet GdRu$_2$Si$_2$~\cite{Khanh2020_squareSkX}, which has substantial weight at the higher harmonic position $\bm{Q}_1+\bm{Q}_2$. This observation excludes the VtX but it is still consistent with the  MX-I or MX-II phases.

\section{Ginzburg-Landau theory}\label{sec:GL}
The universal GL theory ($Q\rightarrow0$) is obtained via the gradient expansion
\begin{equation}
\bm{S}_{\bm{r}+\bm{\delta}}\approx\bm{S}_{\bm{r}}+\left(\bm{\delta}\cdot\nabla\right)\bm{S}_{\bm{r}}+\frac{1}{2}\left(\bm{\delta}\cdot\nabla\right)^{2}\bm{S}_{\bm{r}}+\cdots,
\end{equation}
that leads to the continuum version of Eq.~\eqref{eq:model}:
\begin{align}
\mathcal{H}_{\text{GL}} & =\int\mathrm{d}\bm{r}\bigg\{  -\frac{I_{1}}{2}\left[\left(\partial_{x}\bm{S}_{\bm{r}}\right)^{2}+\left(\partial_{y}\bm{S}_{\bm{r}}\right)^{2}\right] \nonumber \\
 & \quad + \frac{I_{2}}{2}\left(\partial_{x}^{2}\bm{S}_{\bm{r}}+\partial_{y}^{2}\bm{S}_{\bm{r}}\right)^{2} +\frac{\Gamma}{2}\left[\left(\partial_{x}S_{\bm{r}}^{x}\right)^{2}+\left(\partial_{y}S_{\bm{r}}^{y}\right)^{2}\right] \nonumber \\
 & \quad -\left(D - \Gamma\right)\left(S_{\bm{r}}^{z}\right)^{2} -HS_{\bm{r}}^{z} -I_{3} \left(\partial_x^2 \bm{S}_{\bm{r}}\right) \cdot \left(\partial_y^2 \bm{S}_{\bm{r}}\right) \bigg\}.
 \label{eq:GL_dimensionful}
\end{align}

\begin{figure*}[tp]
\centering
\includegraphics[width=1\textwidth]{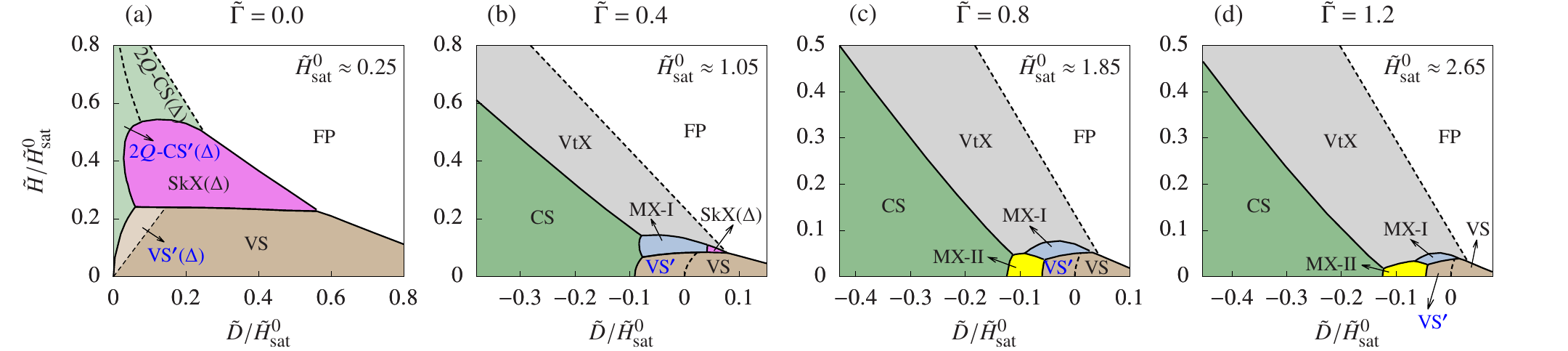}
\caption{Phase diagrams of the GL theory Eq.~\eqref{eq:GL_dimensionless}
with $I_3=0$ at $T=0$. The values of $\tilde{\Gamma}$ are denoted on the top
of each column. {The solid (dashed)  lines  indicate first- (second-) order phase transitions.} \label{fig:phd_GL}}
\end{figure*}

The compass anisotropy and the last term of Eq.~\eqref{eq:GL_dimensionful}   are the only terms of $\mathcal{H}_{\text{GL}}$ that enforce the C$_4$ anisotropy of the original lattice model.
In agreement with the phase diagram
of the microscopic model, the quartic exchange anisotropy penalizes multi-$\bm{Q}$
states for $I_{3}<0$ (bottom row of Fig.~\ref{fig:phd}), while it favors them for $I_{3}>0$
(top row of Fig.~\ref{fig:phd}). This point becomes clearer upon taking the Fourier transform of the $I_3$ term:
\begin{equation}
-I_3 \int \mathrm{d} \bm{q} \left(q_x^2 q_y^2 \right) \bm{S}_{\bm{q}}\cdot \bm{S}_{-\bm{q}}.
\end{equation}

The  characteristic wavelength of the ground state is set by the length scale  $\sqrt{I_2/I_1}$. By adopting this scale as the unit of length $\bm{r}\rightarrow \sqrt{I_1/I_2}\,\bm{r}$, $I_1$ as the unit of energy and $I_1^2/I_2$ as the unit of magnetic field,
the GL energy functional Eq.~\eqref{eq:GL_dimensionful} can be reexpressed in terms of the
dimensionless coupling constants:
\begin{equation}
\tilde{\Gamma}\equiv\frac{\Gamma}{I_{1}}, \quad
\tilde{D}\equiv\frac{I_{2}(D - \Gamma)}{I^2_{1}}, \quad\tilde{H}\equiv\frac{I_{2}H}{I^2_{1}}, \quad\tilde{\mathcal{H}}_{\text{GL}}\equiv\frac{\mathcal{H}_{\text{GL}}}{I_{1}}.
\label{eq:rescaling}
\end{equation}
The resulting GL energy functional reads
\begin{align}
\tilde{\mathcal{H}}_{\text{GL}} & =\int\mathrm{d}\bm{r}\bigg\{-\frac{1}{2}\left[\left(\partial_{x}\bm{S}_{\bm{r}}\right)^{2}+\left(\partial_{y}\bm{S}_{\bm{r}}\right)^{2}\right]+\frac{1}{2}\left(\partial_{x}^{2}\bm{S}_{\bm{r}}+\partial_{y}^{2}\bm{S}_{\bm{r}}\right)^{2}\nonumber \\
& \quad +\frac{\tilde{\Gamma}}{2}\left[\left(\partial_{x}S_{\bm{r}}^{x}\right)^{2}+\left(\partial_{y}S_{\bm{r}}^{y}\right)^{2}\right] - \tilde{D}\left(S_{\bm{r}}^{z}\right)^{2} -\tilde{H}S_{\bm{r}}^{z}\bigg\} \nonumber \\
& \quad  -\frac{I_3}{I_2} \int\mathrm{d}\bm{r}\left(\partial_x^2 \bm{S}_{\bm{r}}\right) \cdot \left(\partial_y^2 \bm{S}_{\bm{r}}\right).
\label{eq:GL_dimensionless}
\end{align}

The $T=0$ variational phase diagrams for the GL theory Eq.~\eqref{eq:GL_dimensionless}
with $\tilde{\Gamma}=\{0,0.4,0.8,1.2\}$ and $I_{3}=0$ (see Fig.~\ref{fig:phd_GL})
are qualitatively similar to
the ones obtained for the microscopic model (Fig.~\ref{fig:phd}).
This fact demonstrates  the universal character  of multi-$\bm{Q}$ orderings.  We can then use the GL functional Eq.~\eqref{eq:GL_dimensionless}
to understand the general principle behind the emergence of the MX and VtX phases.

The first observation is that the ordering wave vectors $\bm{Q}_1$ and $\bm{Q}_2$
of the double-$\bm{Q}$ states remain parallel to the principal $x$ and $y$ directions even for $I_3 >0$. The reason is that  the $\bm{q}$-dependent part of the $\Gamma$ term favors proper screw spirals,
$\bm{S}_{\bm{r}}= ( 0,\cos\left(\bm{Q}_1\cdot\bm{r}\right), \sin\left(\bm{Q}_1\cdot\bm{r}\right) )$ and $\bm{S}_{\bm{r}}= ( \cos\left(\bm{Q}_2\cdot\bm{r}\right), 0,  \sin\left(\bm{Q}_2\cdot\bm{r}\right) )$,
that propagate along these directions. Indeed,  these principal directions become energetically favorable for
$I_3/I_2 < \tilde{\Gamma} (1-\tilde{\Gamma}/8)$.
This observation allows us to understand the stabilization mechanism of the
double-$\bm{Q}$ MX-II phase that is approximately described by the equation
\begin{subequations}
\begin{align}
m_{\bm{r}}^{x} & =\pm a_{1}\sin\left(\bm{Q}_{2}\cdot\bm{r}\right),\\
m_{\bm{r}}^{y} & =\pm a_{1}\sin\left(\bm{Q}_{1}\cdot\bm{r}\right),\\
m_{\bm{r}}^{z} & =a_{0}+a_{2}\cos\left(\bm{Q}_{1}\cdot\bm{r}\right)\cos\left(\bm{Q}_{2}\cdot\bm{r}\right),
\end{align}
\end{subequations}
where $\bm{S}_{\bm{r}} \equiv \bm{m}_{\bm{r}} / |\bm{m}_{\bm{r}}|$.
As is clear from the phase diagram, the competing zero-field phases are the conical (cycloidal) and vertical (proper screw) spiral orderings. A simple evaluation of the energies of these competing phases, which does not include the small higher harmonic components, produces a rough estimate of the stability interval of the MX-II phase at zero field (see Appendix~\ref{sec:stability_MXII} for more details):
\begin{equation}
-\frac{1}{2}+\frac{5}{32}(2-\tilde{\Gamma})^{2}-\frac{I_{3}}{4 I_{2}} \lesssim \tilde{D} \lesssim-\frac{1}{12}+\frac{I_{3}}{6 I_{2}}.
\end{equation}

Note that a positive quartic anisotropy $0<I_3/I_2 < \tilde{\Gamma} (1-\tilde{\Gamma}/8)$ lowers the exchange energy cost of the MX-II phase (the first harmonics of this phase are $\pm {\bm Q}_1 \pm {\bm Q}_2$).  According to Eqs.~\eqref{eq:GL_dimensionful} and \eqref{eq:rescaling}, the $\bm{q}$-independent contribution of the
$\Gamma$ term can turn the single-ion term into an effective easy-plane anisotropy ${\tilde D} <0$.
This explains the suppression of the VS phase in favor of the MX-II phase, which has stronger in-plane (XY) spin components.
The MX-II phase is also more stable than the CS because of the $\bm{q}$-dependent contribution of the $\Gamma$ term  increases the energy cost of the CS phase by an amount proportional to $\tilde{\Gamma} Q^2$ (see Appendix~\ref{sec:stability_MXII} for more details).

For positive $I_3$, the compass anisotropy $\tilde{\Gamma}$ must be strong enough to guarantee that the
ordering wave vectors remain parallel to the principal $x$ and $y$ directions.
These conditions can be naturally fulfilled by tetragonal metallic systems, such as 4$f$-electron materials, with dominant
RKKY interactions and a small Fermi surface. The combination of a small Fermi surface and strong SOC
can naturally lead to $0<I_3/I_2 < \tilde{\Gamma} (1-\tilde{\Gamma}/8)$.

\section{Anomalous Hall effect in 2D}\label{sec:Hall_2D}
The net scalar spin chirality of the  double-$\bm{Q}$ states  MX-I and MX-II creates an effective U(1) gauge flux~\cite{Zhang2020_Berry} for itinerant electrons that are coupled via exchange to the magnetic moments $\bm{S}_i $. The resulting nonzero Berry curvature of the (reconstructed) electronic bands leads to anomalous Hall effect.
Unlike typical realizations of field-induced SkX, the MX-I and MX-II states can {be realized at zero field and} produce a spontaneous topological Hall effect.\footnote{In other centrosymmetric models, SkXs typically only exist for  finite magnetic field and the degeneracy of the SkX solutions with opposite chiralities can be lifted the {\it orbital} coupling to the external  field~\cite{Bluaevskii2008}.}

\begin{figure}[tp]
\centering
\includegraphics[width=1\columnwidth]{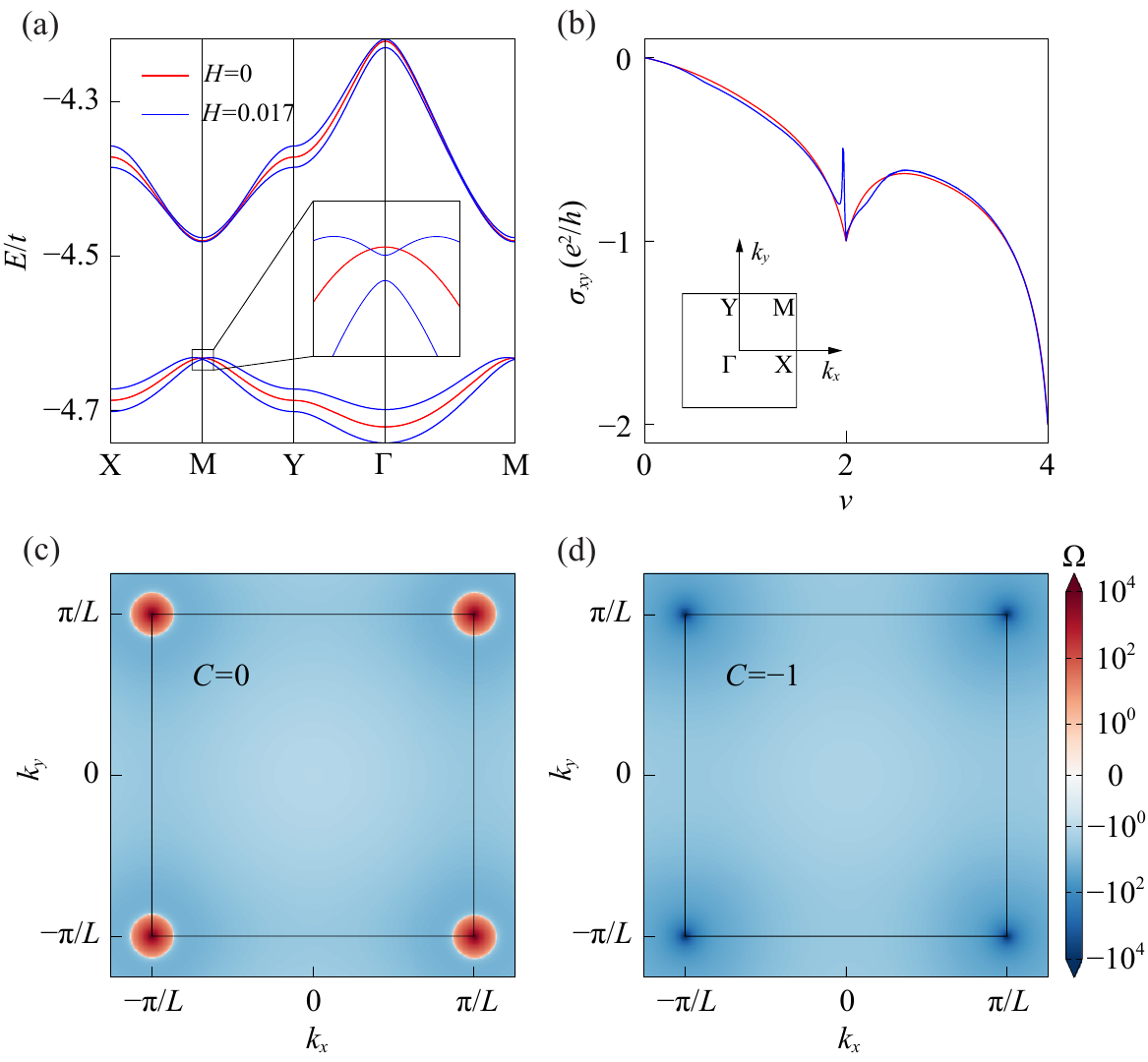}
\caption{Reconstructed band dispersion, transverse conductivity, and Berry curvature of the lowest electronic bands due to coupling to the MX-II state. The MX-II states that we used here are obtained for $J_1=-1$, $J_2=0.35$, $J_3=0.15$, $\Gamma=0.3$, and $D=0.16$ in the $L=8$ variational space (Appendix~\ref{sec:methods}), and $J/t=1$.
(a) Lowest four reconstructed electronic bands due to coupling to $H=0$ MX-II state (red solid curves, doubly degenerate) and MX-II state distorted by an in-plane field along [110] direction with $H=0.017$ (blue solid curves).
(b) Transverse conductivities as a function of filling for the four lowest bands in (a). Here the filling fraction $\nu=n$ indicates the lowest $n$ bands are filled. The inset in (b) shows the first BZ of the magnetic superlattice.
(c), (d) Berry curvatures of the two lowest bands in (a), obtained with the MX-II state distorted by an in-plane field along the [110] direction with $H=0.017$. The black squares enclose the folded first BZ.
\label{fig:anomalous_Hall}}
\end{figure}

\begin{figure*}[tp]
\centering
\includegraphics[width=0.8\textwidth]{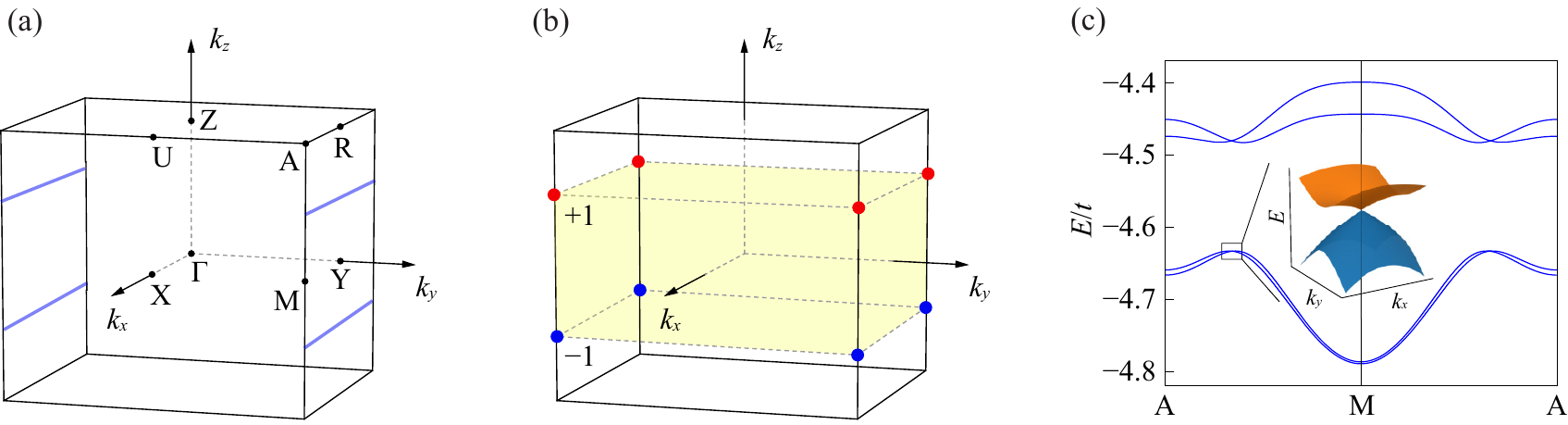}
\caption{Nodal lines and Weyl points generated by coupling vertically stacked MX-II states to itinerant electrons. The MX-II states used in this calculation were obtained with $J_1=-1$, $J_2=0.35$, $J_3=0.15$, $\Gamma=0.3$, and $D=0.16$ in the $L=8$ variational space (Appendix~\ref{sec:methods}).
(a) Nodal lines at
$\bm{k}=(k_x,\pi/L,\pm\pi/2 )$ are generated by distorting the MX-II state with
an in-plane field $H_y$ along the [010] direction, and
a finite SOC in the interlayer hopping $t_z$.
(b) Weyl points marked by red and blue spots with topological charges $\pm 1$, respectively, are generated by further adding intralayer hopping $t_x$ with finite SOC.
(c) Electronic dispersion of the four lowest bands along A$\rightarrow$M$\rightarrow$A direction, where the inset shows a zoomed-in view near the Weyl points. The parameters used in (c) are $J=t$, $H_y=0.012$, $t_z=0.1t$, and $t_x=0.05t$.
\label{fig:Weyl}}
\end{figure*}

This simple phenomenon  can be illustrated by coupling the local moments $\bm{S}_i $ to the spins of itinerant electrons~\cite{Zener51},
\begin{equation}
\mathcal{H}_K=-t\sum_{\langle ij\rangle, \sigma} (c_{i\sigma}^\dagger c_{j\sigma}+\mathrm{H.c.}) + J\sum_{i,\alpha\beta} c_{i\alpha}^\dagger \bm{\sigma}_{\alpha\beta}c_{i\beta} \cdot \bm{S}_i,
\label{eq:kondo}
\end{equation}
where the first term is the nearest-neighbor hopping, and the second term represents the exchange interaction $J$ between the itinerant electrons and the local moments.
By assuming that the effective spin-spin interaction mediated by the conduction electrons is much smaller than the relevant exchange energy scales in Eq.~\eqref{eq:model}, we can still use the magnetic phase diagram obtained
for $\mathcal{H}$ (Fig.~\ref{fig:phd}).
As discussed below, the topological properties of the itinerant electrons are mainly dictated by symmetry, so the choice of nearest-neighbor hopping in Eq.~\eqref{eq:kondo} is a matter of convenience.

For long period ($Q \ll 1$) incommensurate states, the (folded) first Brillouin zone (BZ) includes a large number of reconstructed electronic bands. {The  transverse conductivity depends on the filling fraction. For the purpose of illustration, we will focus on the four lowest energy bands (our results can be easily generalized to include more bands).}
Figure~\ref{fig:anomalous_Hall}(a) shows the reconstructed electronic bands for the MX-II state, which
turn out to be doubly degenerate at  $H=0$.
This degeneracy  is protected by the combined symmetries $\mathcal{O}_1=\mathcal{I} \mathcal{C}_{2 z}$ and $\mathcal{O}_2=\left\{\mathcal{C}_{2 x} \overline{\mathcal{M}}_y|\frac{1}{2},0,0\right\}$ (Appendix~\ref{sec:symmetry}). A finite magnetic field along the $z$ axis breaks $\mathcal{O}_2$ and partially lifts the degeneracy except for the folded BZ boundaries, which are protected by the nonsymmorphic symmetries $\mathcal{O}_3=\left\{\mathcal{T}\mathcal{M}_y|\frac{1}{2},0,0\right\}$ and $\mathcal{O}_4=\left\{\left.\mathcal{T}\mathcal{M}_x\right|0,\frac{1}{2},0\right\}$~\cite{Ramazashvili2008} (Appendix~\ref{sec:symmetry}). The remaining degeneracies can be fully lifted
by applying an in-plane magnetic field along the [110] direction [Fig.~\ref{fig:anomalous_Hall}(a)] that distorts the MX-II state.

The MX-II state produces anomalous Hall conductivity even in absence of magnetic field, implying that the Hall effect is purely of topological origin.
As shown in Fig.~\ref{fig:anomalous_Hall}(b), a value as large as $|\sigma_{xy}|=2e^2/h$ can be achieved by completely filling the four lowest bands.
Figures~\ref{fig:anomalous_Hall}(c) and \ref{fig:anomalous_Hall}(d) show the Berry curvatures of the two lowest bands depicted in Fig.~\ref{fig:anomalous_Hall}(a). The direct field-induced gap between these two bands reaches the minimum value at
the M point of the folded BZ [see inset of Fig.~\ref{fig:anomalous_Hall}(a)] producing a sharp increase of the  Berry curvature [Figs.~\ref{fig:anomalous_Hall}(c) and \ref{fig:anomalous_Hall}(d)] and a sharp peak of $\sigma_{xy}$ around band filling $\nu=2$
[Fig.~\ref{fig:anomalous_Hall}(b)].\footnote{Here the filling $\nu=n$ indicates that the lowest $n$ bands are filled.}
The resulting Chern number of the lower (higher) energy band is $C=0$ ($-1$). Although the Berry curvature of the lowest band is large and positive at the M point, it is negative and small in the rest of the BZ [Fig.~\ref{fig:anomalous_Hall}(c)], leading to a cancellation of the Chern number. Consequently, the Hall conductivity becomes quantized at $\sigma_{xy}=-e^2/h$ for filling fraction $\nu=2$. Similarly, $\sigma_{xy}=-2e^2/h$ for filling fraction $\nu=4$  because  the total Chern number of the four lowest bands is $C=-2$ [see Fig.~\ref{fig:anomalous_Hall}(b)].
Note that the signs of $C$ and $\sigma_{xy}$ are controlled by the sign of scalar chirality of the MX-II state, which can be flipped without energy cost.

\section{Magnetic Weyl semimetal in 3D}\label{sec:Weyl}
Weyl semimetals are realized in crystals with broken time-reversal or inversion symmetry~\cite{Armitage2018_rmp}.
In particular, it was proposed that topological spin textures and Weyl points could affect each other in the same material~\cite{Puphal2020}.
As we demonstrate below, Weyl points can be systematically generated by coupling the conduction electrons
to vertically stacked layers of MX-II states.
The $\mathcal{O}_{n=1,2,3,4}$ symmetries of the 2D MX-II state can be generalized to the 3D MX-II state as $\mathcal{O}_n\overline{\mathcal{M}}_z$ (Appendix~\ref{sec:symmetry}).
In the $H=0$ case, all the bands are doubly degenerate in the whole BZ, and the degeneracy is protected by the combined symmetries $\mathcal{O}_1\overline{\mathcal{M}}_z$ and $\mathcal{O}_2\overline{\mathcal{M}}_z$ (Appendix~\ref{sec:symmetry}).
Weyl points can be generated by partially lifting this double degeneracy, which can be done in different ways.
For instance, the distortion of the spin texture induced by an in-plane field
$H_y$ along the [010] direction
breaks the $\mathcal{O}_1\overline{\mathcal{M}}_z$, $\mathcal{O}_2\overline{\mathcal{M}}_z$, and $\mathcal{O}_3\overline{\mathcal{M}}_z$ symmetries, leaving a nodal surface at the BZ side $\bm{k}=(k_x, \pi/L, k_z)$ protected by $\mathcal{O}_4\overline{\mathcal{M}}_z$,
where $L$ is the lattice constant of the magnetic superlattice. Further inclusion of the SOC in the interlayer hopping $t_z \sum_i \sum_{\alpha\beta} \left( c_{i+\hat{z},\alpha}^\dagger \sigma_{\alpha\beta}^x c_{i\beta} + \mathrm{H.c.} \right)$ gaps out the nodal surfaces except for  two nodal lines at
$\bm{k}=(k_x,\pi/L,\pm\pi/2 )$, where the distance between two nearest-neighbor MX-II layers is set as unity [Fig.~\ref{fig:Weyl}(a)].
The SOC-induced conversion  of the nodal surface into  two nodal lines  can be easily understood by noticing that the Fourier  transform  of the SOC term vanishes at $k_z=\pm\pi/2$ (Appendix~\ref{sec:symmetry}).

Finally, a further inclusion of SOC in the intralayer hopping, $t_x \sum_i \sum_{\alpha\beta} \left( c_{i+\hat{x},\alpha}^\dagger \sigma_{\alpha\beta}^x c_{i\beta} +\mathrm{H.c.} \right)$, lifts the degeneracy of the nodal lines and generates  pairs of Weyl points with opposite chirality in the folded first BZ [Figs.~\ref{fig:Weyl}(b)(c)]. The chirality of a Weyl point is characterized by the topological charge that counts the total Berry curvature flux emitted or absorbed  by the Weyl point~\cite{Wan2011}.  In our system, each pair of vertically aligned Weyl points carry opposite topological charges $C=\pm1$
[Fig.~\ref{fig:Weyl}(b)].
Then 2D energy bands for fixed $k_z$ values between the two Weyl points [highlighted in Fig.~\ref{fig:Weyl}(b)] carry a nonzero Chern number. Thus, the contribution of each pair of Weyl points to the  quantum anomalous Hall effect ($\sigma_{xy}$)  increases with their separation, which can be tuned by changing
the ratio $t_x/t_z$~\cite{Burkov2011}.

\section{Summary and discussion}\label{sec:summary}
Our paper introduces a guiding principle for the stabilization of
topological spin textures in centrosymmetric tetragonal magnets. The magnetic unit cell of these crystals comprises four merons, whose net skyrmion charge can be $\pm2$ (MX-II), $\pm 1$ (MX-I), and
$0$ (VtX). The three phases are double-$\bm{Q}$ orderings obtained from the superposition  of two proper screw (vertical) spirals that propagate along perpendicular directions with a wave number $Q$ dictated by competing exchange interactions (magnetic frustration). The key observation is that the above-mentioned  double-$\bm{Q}$ VS orderings result from the  competition between easy-axis and compass anisotropies. The combination of these two terms can produce a relatively weak \emph{effective easy-plane} anisotropy ${\tilde D}<0$ and a rather strong $\bm{q}$-dependent component of the compass anisotropy ${\tilde \Gamma}$, that penalizes the conical phase.
An effective easy-plane single-ion anisotropy favors the double-$\bm{Q}$ VS orderings
relative to the single-$\bm{Q}$ VS phase because, on average,
the MXs and the VtX have a stronger  in-plane component of the spins than the VS.

It is important to note that these double-$\bm{Q}$ phases could be further stabilized by effective four-spin interactions that
are naturally generated in systems where the interaction between local moments is mediated by conduction electrons~\cite{Ozawa2016,Batista16,Ozawa2017}. In this case, the compass anisotropy alone (without the need of a bare single-ion anisotropy) should be enough to stabilize the MX-I, MX-II, and VtX phases. Since these conditions can be naturally fulfilled by $f$-electron mangets, our results indicate that tetragonal  $f$-electron materials with relatively weak single-ion aniotropy are natural candidates to host these phases. This hypothesis is supported by the recent experimental results in the centrosymmetric tetragonal material GdRu$_2$Si$_2$~\cite{Khanh2020_squareSkX}.

{While we have focused on the $T=0$ limit, all of the reported phases break discrete symmetries, which should not be immediately restored by thermal fluctuations.
The finite temperature phase diagram and its dependence on dimensionality are left for future studies based on Monte Carlo simulations~\cite{Okubo12}.}

Finally, the  effective U(1) gauge flux~\cite{Zhang2020_Berry} generated by the MXs produces  anomalous Hall effect in metallic systems even at zero field. Moreover, a symmetry analysis reveals that the reconstructed electronic bands can naturally include  Weyl points for 3D versions of the MX phases, that lead to a magnetic Weyl semimetal for certain filling fractions.

{\it Note added in proof}. Recently, we became aware of a study on an effective spin model with long-range anisotropic interactions defined in the momentum space, that reports some of the multi-$\bm{Q}$ spin structures in this work, including the MX-I and VtX phases~\cite{Hayami2021_square}.

\begin{acknowledgements}
Z.W. and C.D.B. were supported by funding from the Lincoln Chair of Excellence in Physics.
During the writing of this paper, Z.W. was supported by the U.S. Department of Energy through the University of Minnesota Center for Quantum Materials, under Award No. DE-SC-0016371.
The work at Los Alamos National Laboratory (LANL) was carried out
under the auspices of the U.S. DOE NNSA under Contract No.
89233218CNA000001 through the LDRD Program, and was supported by the
Center for Nonlinear Studies at LANL.
This research used resources of the Oak Ridge Leadership Computing
Facility at the Oak Ridge National Laboratory, which is supported
by the Office of Science of the U.S. Department of Energy under Contract
No. DE-AC05-00OR22725.
\end{acknowledgements}

\appendix

\section{Variational Methods}\label{sec:methods}
The variational methods used in this paper  follow closely Ref.~\cite{Wang2020_RKKYskx} except for some fine details.

We start by assuming that the ground-state magnetic super-cell contains
$L\times L$ spins spanned by the basis \{$L\bm{a}_{1}$, $L\bm{a}_{2}$\},
where $\bm{a}_{1}$ and $\bm{a}_{2}$ are the basis of the square
lattice. In other words, the ground-state ordering wave number $Q$
is fixed by $Q=l|\bm{b}_{1}|/L$, where $l$ is another integer
satisfying $l<L$. We first try several integer values of $L$, which
is given by
\begin{equation}
L\approx2\pi/Q_{0},\quad2\pi/Q_{0}\pm1,\quad2\pi/Q_{0}\pm2\ldots,
\end{equation}
where $Q_{0}$ is the wave number that minimizes Eq.~\eqref{eq:exchange_kspace}.
Clearly, if the true ground state is incommensurate to the choices
of $L$ (typically ranges from 4 to 10 for the parameters considered in this work), a systematic error is introduced
in the energy calculation. To resolve this issue, we will relax this
variational constraint later.

The spin states are described by $2\times L\times L$ variational
parameters \{$\theta_{\bm{r}}$, $\phi_{\bm{r}}$\}
that parametrize the spin space with $|\bm{S}_{\bm{r}}|=1$.
The total energy density of Eq.~\eqref{eq:model}
is then numerically minimized in this variational space~\cite{nlopt}.
Finally, for any given parameter set \{$J_{1}$, $J_{2}$, $J_{3}$,
$\Gamma$, $D$, $H$\}, we pick the lowest energy solution among
the different choices of $L$.

Such a variational calculation is unbiased, in the sense that we are
not making any assumption of the spin structure, except for the size
of the underlying magnetic unit cell. However, the phase boundaries
will not be very accurate since the ordering wave number $Q$ is generally
incommensurate. To further improve the phase boundaries, we will perform
another calculation where $Q$ is included as a variational parameter.

To prepare for the refined variational calculation, we first perform
a Fourier analysis of the phases discovered by the unbiased variational
method and parametrize them by a few dominant Fourier components (Appendix~\ref{sec:fourier}).
With such parametrization, we can relax the magnetic supercell assumption,
and include the ordering wave number as a variational parameter. For
each point in the phase diagrams, we compare the optimized energies of different
ansatze and pick the lowest one.

The energy minimization of different ansatze are performed on a finite
$512\times512$ square lattice for the microscopic model~\cite{nlopt}.
For the GL theory, we obtain the $T=0$ phase diagrams by discretizing
the real space continuum and replacing derivatives by finite difference. Then,
we optimize different variational ansatze and find the lowest energy
solution. The linear dimension of the real space continuum is set to 200,
and we typically discretize it by a $200\times200$ mesh. Note that
the phase boundary between VS$^{\bm{\prime}}$($\Delta$) and VS states
[Fig.~\ref{fig:phd_GL}(a)] requires a mesh as fine as $1600\times1600$
for good convergence.

\section{Ordering wave vectors of the Heisenberg exchanges}\label{sec:appendix_Q}
\begin{figure}[tp]
\centering
\includegraphics[width=\columnwidth]{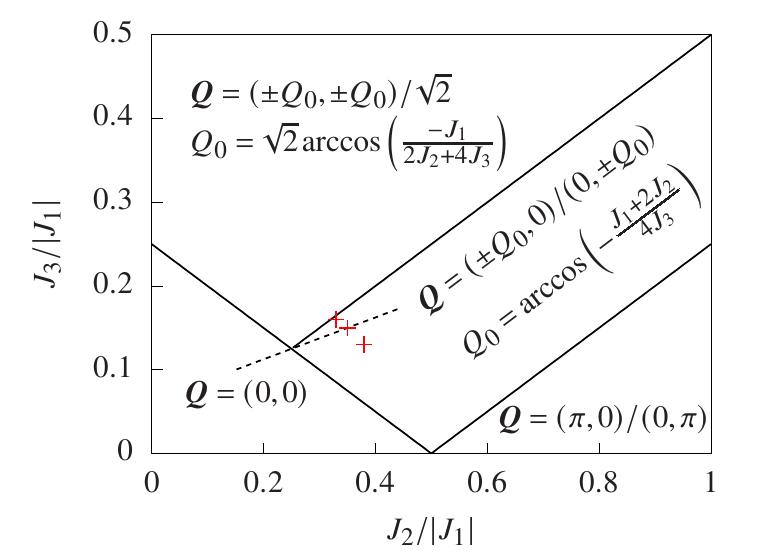}
\caption{Phase diagram showing the positions of the ordering wave vectors
for the square lattice Heisenberg model with $J_{1}<0$ (FM) at $T=0$.
The symbols correspond to the values of \{$J_{1}$, $J_{2}$, $J_{3}$\}
used in Fig.~\ref{fig:phd} in the main text. The dashed line corresponds to $I_3=0$.}
\label{fig:phd_wavevector}
\end{figure}

The ordering wave vectors of the square lattice Heisenberg model (no
anisotropy or magnetic field) can be determined by minimizing the
exchange interaction Eq.~\eqref{eq:exchange_kspace}.
The results for $J_{1}<0$ (FM) are presented in Fig.~\ref{fig:phd_wavevector}.
Note that this analysis is no longer valid in the presence of magnetic field (Zeeman) and spin anisotropies because the lowest energy state is not necessarily a single-$\bm{Q}$ spiral ordering. Consequently, $\bm{Q}$ must be treated as an additional
variational parameter to determine the lowest energy state (Appendix~\ref{sec:methods}).

Depending on the sign of $2J_3-J_2$,
the incommensurate ordering wave vectors are located on different principle axes (Fig.~\ref{fig:phd_wavevector}):

(1) For $2J_3 < J_2$, $\bm{Q} = (\pm Q_0,\,0)$ or $(0,\, \pm Q_0)$, where $Q_0 = \arccos \left( -\frac{J_1+2J_2}{4J_3} \right)$.

(2) For $2J_3>J_2$, $\bm{Q}=(\pm Q_0, \, \pm Q_0 )/\sqrt{2}$, where $Q_0=\sqrt{2}\arccos \left( \frac{-J_1}{2J_2 + 4J_3} \right)$.

In Fig.~\ref{fig:phd} in the main text, we take three points in the region with
$\bm{Q}=(\pm Q_{0},0)/(0,\pm Q_{0})$, all of which gives roughly
$Q_{0}\approx\frac{2\pi}{6}$ (see Fig.~\ref{fig:phd_wavevector}).

In the long wavelength limit, we can expand Eq.~\eqref{eq:exchange_kspace},
\begin{equation}
J(\bm{q}) \simeq I_0 - \frac{I_1}{2} q^2 + \frac{I_2}{2} q^4 - I_3 q^2_x q^2_y,
\end{equation}
with $I_0=2(J_1+J_2+J_3)$ and
\begin{equation}
I_{1}  =J_{1}+2J_{2}+4J_{3}, \;\;
I_{2}  =\frac{J_{1}+2J_{2}+16J_{3}}{12}, \;\;
I_{3}  =I_2 - \frac{J_{2}}{2}.
\end{equation}

In this limit, the direction of the ordering wave vector $\bm{Q}$ is determined by the sign of the quartic anisotropy $I_3$ (see dashed line in Fig.~\ref{fig:phd_wavevector}):

(1) For $I_3<0$, $\bm{Q} = (\pm Q_0,\,0)$ or $(0,\, \pm Q_0)$, where $Q_0 = \sqrt{I_1/2I_2}$.

(2) For $I_3>0$, $\bm{Q}=(\pm Q_0, \, \pm Q_0 )/\sqrt{2}$, where $Q_0 = \sqrt{I_1/(2I_2-I_3)}$.

\begin{figure*}[tbp]
\centering
\includegraphics[width=\textwidth]{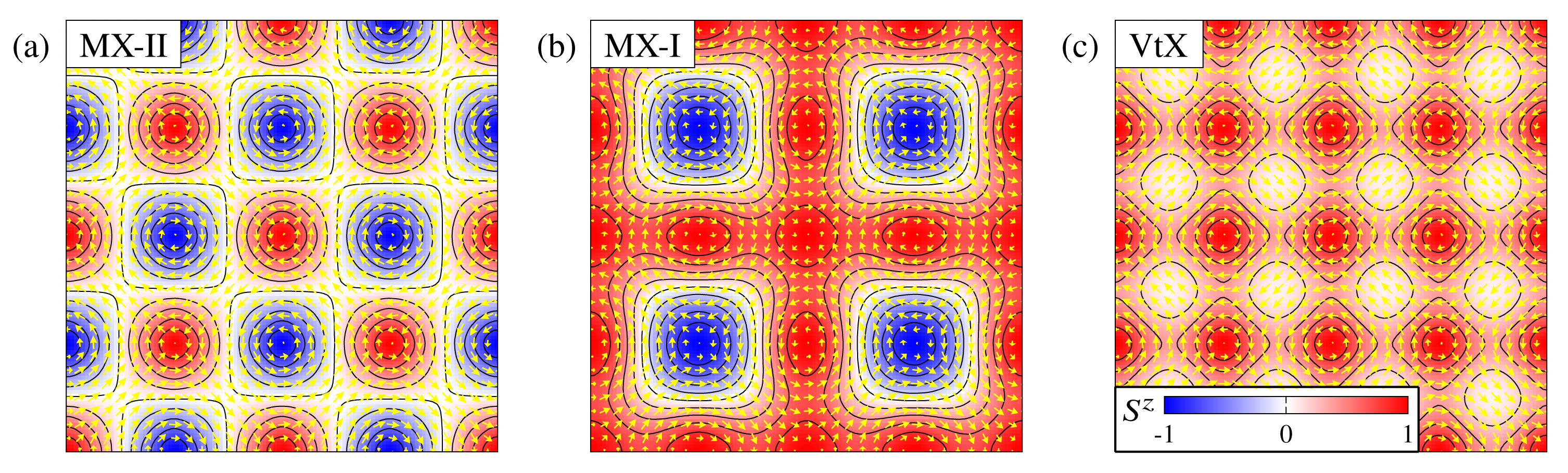}
\caption{Vector-contour plots of the MX-II, MX-I, and VtX states {with the parametrization {given in} Eq.~\eqref{eq:MX_universal}}. The
arrows represent the $xy$ components of the spins, and the background colors represent
the $z$ component of the spins. The contours are taken at linear incremental
level with $\Delta S^{z}=0.2$. {(a) The parameters used in MX-II are $a_0=0.025$, $a_1=1$, $a_4=0.016$, $a_7=0.689$, $a_9=0.056$, and $a_2=a_3=a_5=a_6=a_8=0$. (b) The parameters used in MX-I are $a_0=0.288$, $a_1=1$, $a_2=-0.345$, $a_3=-0.173$, $a_4=0.067$, $a_5=0.48$, $a_6=0.044$, $a_7=-0.762$, $a_8=0.081$, and $a_9=0.024$. (c) The parameters used in VtX are $a_0=0.395$, $a_1=1$, $a_4=-0.011$, $a_6=0.162$, $a_9=0.03$, and $a_2=a_3=a_5=a_7=a_8=0$.}  }
\label{fig:contour}
\end{figure*}

\section{Skyrmion charge of MX-I, MX-II, and VtX states}\label{sec:chargeQ}
The skyrmion charge is defined by
\begin{equation}
 n_{\text{sk}}=\frac{1}{4\pi}\int \mathrm{d}\bm{r}\, \bm{S}_{\bm{r}} \cdot \left( \partial_x\bm{S}_{\bm{r}} \times \partial_y\bm{S}_{\bm{r}} \right),
\end{equation}
which corresponds to the area spanned by the unit vector field $\bm{S}_{\bm{r}}$ (in
units of the area $4 \pi$ of the unit sphere) when the two-dimensional (2D) real space $\mathbb{R}^2\cup \{\infty\}$ is mapped to $S^2$ via stereographic projection. A single skyrmion  wraps the sphere only once and yields $n_{\text{sk}}=\pm1$ (the sign is defined by the sign of scalar chirality). A meron corresponds to a half covering,
which yields $n_{\text{sk}}=\pm1/2$, while the skyrmion charge of a
vortex is generally an arbitrary  number different from 1 or  1/2.

Figure~\ref{fig:contour} shows the vector-contour plots of the MX-II, MX-I, and VtX states.
It is clear that the MX-II state contains
four merons in each magnetic unit cell.
The skyrmion charge for a magnetic unit cell can be read out directly from  Fig.~\ref{fig:contour}(a):
$n_{\text{sk}}=-\frac{1}{2}\times4=-2$. Note that the other choice
of $n_{\text{sk}}=2$ has the same energy for the model Eq.~\eqref{eq:model}
considered in the main text, which can be obtained by simply changing
the sign of $S_{\bm{r}}^{y}$ at every site.
The MX-I state contains
four vortices (or, equivalently, four overlapping merons) in each magnetic unit cell [Fig.~\ref{fig:contour}(b)].
Denote the skyrmion charge of the largest vortex ($S_{\bm{r}}^{z}<0$
in the center) as $\alpha$ ($1/2<\alpha<1$), then the two vortices
centered at the elliptical contours have skyrmion charge roughly $1-\alpha$
each, and the last vortex has skyrmion charge $-(1-\alpha)$. Thus,
the net skyrmion chage of MX-I is $n_{\text{sk}}=\alpha+2\times(1-\alpha)-(1-\alpha)=1$.
As is the same in the MX-II case, the $n_{\text{sk}}=\pm1$ solutions have the
same energy in MX-I. The VtX state has four vortices in the magnetic
unit cell, whose skyrmion charges cancel out with each other, giving $n_{\text{sk}}=0$
[Fig.~\ref{fig:contour}(c)].

\section{Fourier analysis of different states}\label{sec:fourier}
The states that appear in this paper can be parametrized by a few dominant Fourier components. We first focus on the single-$\bm{Q}$ and double-$\bm{Q}$ states, which involve
at most two perpendicular wave vectors,
\begin{equation}
\bm{Q}_{1}=(Q,\,0),\quad\bm{Q}_{2}=(0,\,Q),
\end{equation}
and $0<Q<\pi$.

The normalized spin configurations $\bm{S}_{\bm{r}}\equiv\bm{m}_{\bm{r}}/|\bm{m}_{\bm{r}}|$
of the VS and VS$^{\bm{\prime}}$ states can be parametrized by
\begin{subequations}
\begin{align}
m_{\bm{r}-\bm{r}_{0}}^{x} & =a_{1}\sin\left(\bm{Q}_{2}\cdot\bm{r}\right)+a_{2}\sin\left(2\bm{Q}_{2}\cdot\bm{r}\right) \nonumber \\
&\quad+a_{3}\sin\left(3\bm{Q}_{2}\cdot\bm{r}\right),\\
m_{\bm{r}-\bm{r}_{0}}^{y} & =a_{4}\sin\left(\bm{Q}_{1}\cdot\bm{r}\right),\\
m_{\bm{r}-\bm{r}_{0}}^{z} & =a_{0}+a_{5}\cos\left(\bm{Q}_{2}\cdot\bm{r}\right)+a_{6}\cos\left(2\bm{Q}_{2}\cdot\bm{r}\right) \nonumber \\
&\quad+a_{7}\cos\left(3\bm{Q}_{2}\cdot\bm{r}\right).
\end{align}
\end{subequations}
The difference between the VS and VS$^{\bm{\prime}}$
states is that $a_{4}=0$ in the VS while it is nonzero in the VS$^{\bm{\prime}}$.

The normalized spin configurations $\bm{S}_{\bm{r}}\equiv\bm{m}_{\bm{r}}/|\bm{m}_{\bm{r}}|$
of the CS state can be parametrized by\begin{subequations}
\begin{align}
m_{\bm{r}-\bm{r}_{0}}^{x} & =a_{1}\cos\left(\bm{Q}_{1}\cdot\bm{r}\right)+a_{2}\cos\left(3\bm{Q}_{1}\cdot\bm{r}\right),\\
m_{\bm{r}-\bm{r}_{0}}^{y} & =a_{3}\sin\left(\bm{Q}_{1}\cdot\bm{r}\right)+a_{4}\sin\left(3\bm{Q}_{1}\cdot\bm{r}\right),\\
m_{\bm{r}-\bm{r}_{0}}^{z} & =a_{0}+a_{5}\cos\left(2\bm{Q}_{1}\cdot\bm{r}\right).
\end{align}
\end{subequations}

The normalized spin configurations $\bm{S}_{\bm{r}}\equiv\bm{m}_{\bm{r}}/|\bm{m}_{\bm{r}}|$
of the MX-I, MX-II, and VtX states can be universally parametrized by
\begin{widetext}
\begin{subequations}
\begin{align}
m_{\bm{r}-\bm{r}_{0}}^{x} & =\pm\sin\left(\bm{Q}_{2}\cdot\bm{r}\right)\left[a_{1}+a_{2}\cos\left(\bm{Q}_{1}\cdot\bm{r}\right)+a_{3}\cos\left(\bm{Q}_{2}\cdot\bm{r}\right)
+a_{4}\cos\left(2\bm{Q}_{1}\cdot\bm{r}\right)+a_{9}\left(1+2\cos\left(2\bm{Q}_{2}\cdot\bm{r}\right)\right)\right],\\
m_{\bm{r}-\bm{r}_{0}}^{y} & =\pm\sin\left(\bm{Q}_{1}\cdot\bm{r}\right)\left[a_{1}+a_{2}\cos\left(\bm{Q}_{2}\cdot\bm{r}\right)+a_{3}\cos\left(\bm{Q}_{1}\cdot\bm{r}\right)
+a_{4}\cos\left(2\bm{Q}_{2}\cdot\bm{r}\right)+a_{9}\left(1+2\cos\left(2\bm{Q}_{1}\cdot\bm{r}\right)\right)\right],\\
m_{\bm{r}-\bm{r}_{0}}^{z} & =a_{0}+a_{5}\left[\cos\left(\bm{Q}_{1}\cdot\bm{r}\right)+\cos\left(\bm{Q}_{2}\cdot\bm{r}\right)\right]
+a_{6}\left[\cos\left(2\bm{Q}_{1}\cdot\bm{r}\right)+\cos\left(2\bm{Q}_{2}\cdot\bm{r}\right)\right] \nonumber \\
&\quad+a_{7}\cos\left(\bm{Q}_{1}\cdot\bm{r}\right)\cos\left(\bm{Q}_{2}\cdot\bm{r}\right)
+a_{8}\left[\cos\left(2\bm{Q}_{1}\cdot\bm{r}\right)\cos\left(\bm{Q}_{2}\cdot\bm{r}\right)
+\cos\left(2\bm{Q}_{2}\cdot\bm{r}\right)\cos\left(\bm{Q}_{1}\cdot\bm{r}\right)\right],
\end{align}\label{eq:MX_universal}
\end{subequations}
\end{widetext}
where the $\pm$ choice controls the sign of the
skyrmion charge of each meron and vortex.

We continue with the parametrization of the triple-$\bm{Q}$ states, which
are known to exist in the long wavelength limit from the GL theory~\cite{Lin2016_skyrmion}.
Note that these states are not favored by the spatial anisotropy of the square lattice.

The ordering wave vectors \{$\bm{Q}_{1}$, $\bm{Q}_{2}$, $\bm{Q}_{3}$\} are
parametrized by a single parameter in the continuum limit  because $|\bm{Q}_{1}|=|\bm{Q}_{2}|=|\bm{Q}_{3}|$
and their directions differ by $\pm120^\circ$ rotations.
The following parametrization:
\begin{equation}
\begin{split}
\bm{Q}_{1} &=\left(Q_a \cos\alpha,\, -Q_a \sin\alpha \right), \\
\bm{Q}_{2} &=(0,\, Q_{b}),\\
\bm{Q}_{3} &=-\left(Q_{a}\cos\alpha,\, Q_{a}\sin\alpha \right),
\end{split}
\end{equation}
includes the three ordering wave vectors in the continuum limit of the theory (very weak lattice anisotropy) as well as
the corrections by the square lattice anisotropy.
The optimal values from the minimization are found to be $Q_{a}\approx Q_{b}$ and $\alpha\approx\frac{\pi}{6}$ on a
finite square lattice (the approximated relations become exact in  the thermodynamic limit).
Then, the parametrization of the triple-$\bm{Q}$
states follow the ones in Refs.~\cite{Leonov2015,Lin2016_skyrmion,Wang2020_RKKYskx}.

The normalized spin configurations $\bm{S}_{\bm{r}}\equiv\bm{m}_{\bm{r}}/|\bm{m}_{\bm{r}}|$
of the VS$^{\bm{\prime}}$($\Delta$) state can be parametrized as
\begin{subequations}
\begin{align}
m_{\bm{r}-\bm{r}_{0}}^{x} & =a_{1}\cos\phi\sin\left(\bm{Q}_{2}\cdot\bm{r}\right) \nonumber \\
&\quad +a_{2}\sin\phi\left[\cos\left(\bm{Q}_{1}\cdot\bm{r}\right)+\cos\left(\bm{Q}_{3}\cdot\bm{r}\right)\right],\\
m_{\bm{r}-\bm{r}_{0}}^{y} & =a_{1}\sin\phi\sin\left(\bm{Q}_{2}\cdot\bm{r}\right) \nonumber \\
&\quad -a_{2}\cos\phi\left[\cos\left(\bm{Q}_{1}\cdot\bm{r}\right)+\cos\left(\bm{Q}_{3}\cdot\bm{r}\right)\right],\\
m_{\bm{r}-\bm{r}_{0}}^{z} & =a_{0}-a_{1}\cos\left(\bm{Q}_{2}\cdot\bm{r}\right).
\end{align}
\end{subequations}

The normalized spin configurations $\bm{S}_{\bm{r}}\equiv\bm{m}_{\bm{r}}/|\bm{m}_{\bm{r}}|$
of the 2$\bm{Q}$-CS($\Delta$) and 2$\bm{Q}$-CS$^{\bm{\prime}}$($\Delta$)
states can be parametrized as\begin{subequations}
\begin{align}
m_{\bm{r}-\bm{r}_{0}}^{x} & =a_{1}\sin\left(\bm{Q}_{1}\cdot\bm{r}\right)+a_{2}\sin\left(\bm{Q}_{3}\cdot\bm{r}\right),\\
m_{\bm{r}-\bm{r}_{0}}^{y} & =a_{1}\cos\left(\bm{Q}_{1}\cdot\bm{r}\right)-a_{2}\cos\left(\bm{Q}_{3}\cdot\bm{r}\right),\\
m_{\bm{r}-\bm{r}_{0}}^{z} & =a_{0}+a_{3}\cos\left(\bm{Q}_{2}\cdot\bm{r}\right),
\end{align}
\end{subequations}where $a_{1}=a_{2}$ in the 2$\bm{Q}$-CS($\Delta$)
state, while $a_{1}\neq a_{2}$ in the 2$\bm{Q}$-CS$^{\bm{\prime}}$($\Delta$)
state.

The normalized spin configurations $\bm{S}_{\bm{r}}\equiv\bm{m}_{\bm{r}}/|\bm{m}_{\bm{r}}|$
of the SkX($\Delta$) state can be parametrized as
\begin{widetext}
\begin{subequations}
\begin{align}
m_{\bm{r}-\bm{r}_{0}}^{x} & =a_{1}\left[\cos\left(\varphi\right)\sin\left(\bm{Q}_{1}\cdot\bm{r}\right)+\cos\left(\varphi+\kappa\frac{2\pi}{3}\right)\sin\left(\bm{Q}_{2}\cdot\bm{r}\right)
+\cos\left(\varphi+\kappa\frac{4\pi}{3}\right)\sin\left(\bm{Q}_{3}\cdot\bm{r}\right)\right],\\
m_{\bm{r}-\bm{r}_{0}}^{y} & =a_{1}\left[\sin\left(\varphi\right)\sin\left(\bm{Q}_{1}\cdot\bm{r}\right)+\sin\left(\varphi+\kappa\frac{2\pi}{3}\right)\sin\left(\bm{Q}_{2}\cdot\bm{r}\right)
+\sin\left(\varphi+\kappa\frac{4\pi}{3}\right)\sin\left(\bm{Q}_{3}\cdot\bm{r}\right)\right],\\
m_{\bm{r}-\bm{r}_{0}}^{z} & =a_{0}-a_{2}\left[\cos\left(\bm{Q}_{1}\cdot\bm{r}\right)+\cos\left(\bm{Q}_{2}\cdot\bm{r}\right)+\cos\left(\bm{Q}_{3}\cdot\bm{r}\right)\right],
\end{align}
\end{subequations}
\end{widetext}
where $\kappa=\pm1$ controls the sign of the skyrmion
charge, and $\varphi$ controls the helicity. For simplicity, we have
neglected the phases \{$\theta_{1}$, $\theta_{2}$\} in Ref.~\cite{Wang2020_RKKYskx},
which are negligible when $|\bm{Q}_{\nu}|\ll 1$.

\section{Stability estimate of the MX-II phase}\label{sec:stability_MXII}
To understand the stabilization mechanism  of the MX-II phase at zero field,
it is helpful to obtain rough estimates of the energies of the CS, MX-II, and VS states. Here, we work in the long wavelength limit with the rescaled GL functional.

The CS state at zero field can be approximated by
\begin{equation}
\bm{S}_{\bm{r}} = \left( \cos \left(\bm{Q}_1\cdot \bm{r}\right), \, \sin \left(\bm{Q}_1\cdot \bm{r}\right), \, 0 \right),
\end{equation}
whose energy density is
\begin{equation}
e_\text{CS} =  -\frac{Q^2}{2} + \frac{Q^4}{2} + \frac{\tilde{\Gamma}}{4}Q^2
= -\frac{(2-\tilde{\Gamma})^2}{32},
\end{equation}
where the optimal $Q=\frac{\sqrt{2-\tilde{\Gamma}}}{2}$.

The VS state at zero field can be approximated by
\begin{equation}
\bm{S}_{\bm{r}} = \left( \sin \left(\bm{Q}_2\cdot\bm{r}\right),\,0,\,\cos \left( \bm{Q}_2\cdot\bm{r}\right) \right),
\end{equation}
whose energy density is
\begin{equation}
e_\text{VS} =   -\frac{Q^2}{2} + \frac{Q^4}{2} - \frac{\tilde{D}}{2}
=  - \frac{1}{8}  - \frac{\tilde{D}}{2} ,
\end{equation}
where the optimal $Q=\frac{1}{\sqrt{2}}$.

The MX-II state at zero field is approximated by
\begin{subequations}
\begin{align}
m_{\bm{r}}^{x}	&= \pm a_1 \sin \left(\bm{Q}_{2}\cdot\bm{r} \right), \\
m_{\bm{r}}^{y}  &= \pm a_1\sin \left(\bm{Q}_{1}\cdot\bm{r} \right), \\
m_{\bm{r}}^{z}	&= a_2 \cos \left(\bm{Q}_{1}\cdot\bm{r}\right) \cos \left(\bm{Q}_{2}\cdot\bm{r} \right),
\end{align}
\end{subequations}
where $\bm{S}_{\bm{r}} = \bm{m}_{\bm{r}}/ | \bm{m}_{\bm{r}} |$.

For $a_1=a_2$, we obtain
\begin{subequations}
\begin{align}
S_{\bm{q}}^{x} &= \pm \sqrt{\frac{4}{5}}\pi\frac{\delta(\bm{q}+\bm{Q}_{2})-\delta(\bm{q}-\bm{Q}_{2})}{\iu}, \\
S_{\bm{q}}^{y} &= \pm \sqrt{\frac{4}{5}}\pi\frac{\delta(\bm{q}+\bm{Q}_{1})-\delta(\bm{q}-\bm{Q}_{1})}{\iu},\\
S_{\bm{q}}^{z} &= \sqrt{\frac{1}{5}}\pi \Big[ \delta(\bm{q}+\bm{Q}_{1}+\bm{Q}_{2})+\delta(\bm{q}-\bm{Q}_{1}-\bm{Q}_{2}) \nonumber \\ &\quad \quad\quad +\delta(\bm{q}+\bm{Q}_{1}-\bm{Q}_{2})+\delta(\bm{q}-\bm{Q}_{1}+\bm{Q}_{2}) \Big],
\end{align}
\end{subequations}
after applying the sum rule in momentum space (note that we are  neglecting higher harmonics).
The energy density of this state is
\begin{align}
e_\text{MX-II} &=  -\frac{3}{5}Q^{2}+\frac{2}{5}\left(2-\frac{I_{3}}{2I_{2}}\right)Q^{4}-\frac{1}{5}\tilde{D} \nonumber \\
&=-\frac{1}{10}-\frac{I_{3}}{20I_{2}}-\frac{1}{5}\tilde{D},
\end{align}
where we have fixed $Q=\frac{1}{\sqrt{2}}$.

A comparison of the energies leads to an estimate of the stability condition for the MX-II state at zero field:
\begin{equation}
-\frac{1}{2}+\frac{5}{32}(2-\tilde{\Gamma})^{2}-\frac{I_{3}}{4 I_{2}} \lesssim \tilde{D} \lesssim-\frac{1}{12}+\frac{I_{3}}{6 I_{2}}.
\end{equation}

\section{Symmetry analysis of band degeneracy}\label{sec:symmetry}
In this Appendix, we analyze the symmetries of the MX-II state which put restrictions on the electronic band structures of Eq.~\eqref{eq:kondo} in the main text.

\begin{figure*}[tp]
\centering
\includegraphics[width=1\textwidth]{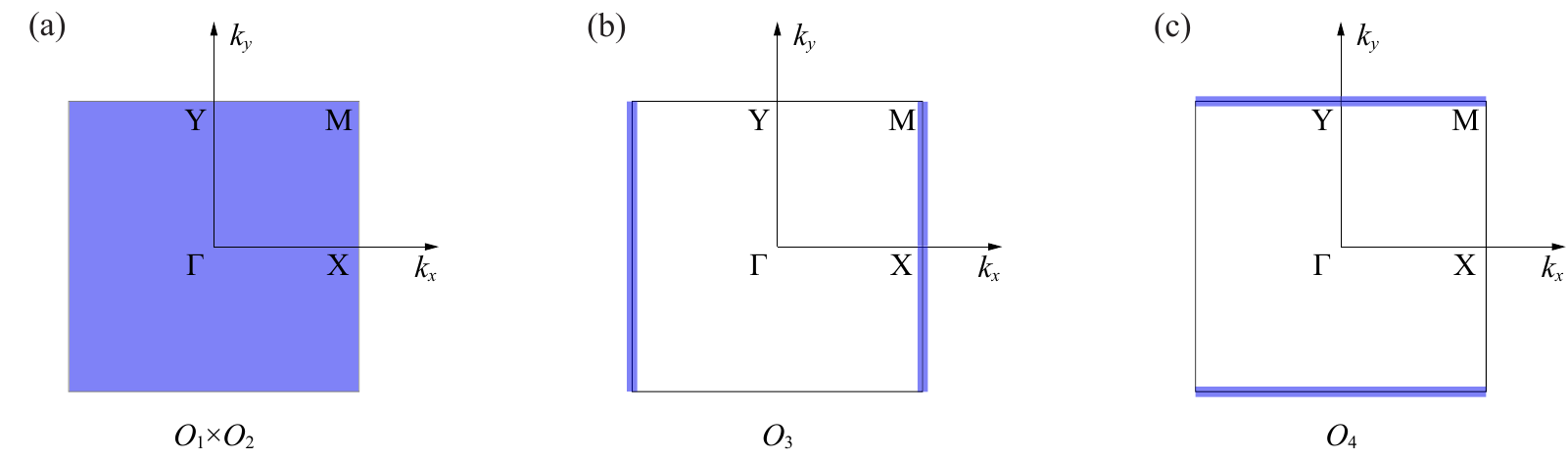}
\caption{Double degeneracy of the electronic bands coupled to 2D MX-II state.
Blue shades mark the double degeneracy protected by the corresponding symmetries.}
\label{fig:symmetry}
\end{figure*}

\begin{figure*}
\centering
\includegraphics[width=1\textwidth]{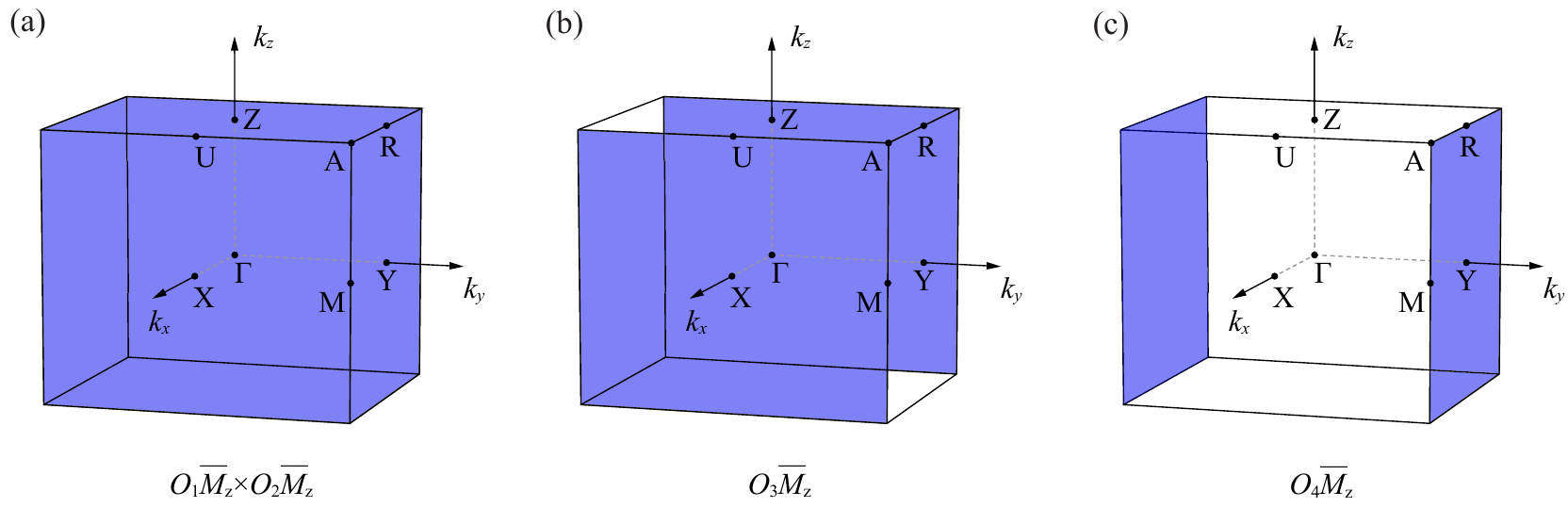}
\caption{Double degeneracy of the electronic bands coupled to vertically stacked MX-II states.
Blue shades mark the double degeneracy protected by the corresponding symmetries.}
\label{fig:symmetry_3d}
\end{figure*}

The zero-field MX-II state has  four important symmetries:

(1) $\mathcal{O}_1=\mathcal{I} \mathcal{C}_{2 z}$;

(2) $\mathcal{O}_2=\left\{\mathcal{C}_{2 x} \overline{\mathcal{M}}_y|\frac{1}{2},0,0\right\}$;

(3) $\mathcal{O}_3=\left\{\mathcal{T}\mathcal{M}_y|\frac{1}{2},0,0\right\}$;

(4) $\mathcal{O}_4=\left\{\left.\mathcal{T}\mathcal{M}_x\right|0,\frac{1}{2},0\right\}$.

Here $\mathcal{O}_2$, $\mathcal{O}_3$, and $\mathcal{O}_4$ are nonsymmorphic symmetries with a glide reflection operation. $\overline{\mathcal{M}}_y$ is the space reflection operator under which $y\rightarrow -y$, while $\mathcal{M}_y$ is the reflection operator acting on both space coordinate and spin as $y\rightarrow -y$, $S^x\rightarrow -S^x$, and $S^z\rightarrow -S^z$. $\mathcal{I}$ is the inversion operator and $\mathcal{T}$ is the time-reversal operator. $C_{2x}$ and $C_{2z}$ are two-fold rotation operators with the rotation axis along the $x$ and $z$ directions, respectively. $\{\frac{1}{2},0,0\}$ represents a translation along the $x$ direction by half a lattice constant of the magnetic superlattice.

As shown in the main text, the MX-II state is stable even in the absence of  magnetic field. The four symmetries are present at zero field, resulting in double degeneracy of the electronic bands over the full BZ [see Fig.~\ref{fig:symmetry}(a)]. To prove this point, consider the symmetries $\mathcal{O}_1$ and $\mathcal{O}_2$. Because $\mathcal{O}_1^2=-1$ and $[\mathcal{O}_1, H(\bm{k})]=0$, we can label the Bloch eigenfunctions of the Hamiltonian  given in Eq.~\eqref{eq:kondo} by the eigenvalues $\pm \iu$ of the $\mathcal{O}_1$:
\begin{equation}
H(\bm{k})|\psi _{\pm }(\bm{k})\rangle =E_{\pm }(\bm{k})|\psi _{\pm }(\bm{k})\rangle,
\end{equation}
and
\begin{equation}
\mathcal{O}_1|\psi _{\pm }(\bm{k})\rangle =\pm \iu|\psi _{\pm }(\bm{k})\rangle.
\end{equation}
Meanwhile, $[\mathcal{O}_2, H(\bm{k})]=0$ gives
\begin{equation}
H(\bm{k})\mathcal{O}_2|\psi _{\pm }(\bm{k})\rangle =E_{\pm }(\bm{k})\mathcal{O}_2|\psi _{\pm }(\bm{k})\rangle.
\end{equation}

We then have two sets of eigenstates $\ket{\psi_\pm(\bm{k})}$ and $\mathcal{O}_2\ket{\psi_\pm(\bm{k})}$ that have the same eigenvalues $E_\pm(\bm{k})$. Furthermore, since $\mathcal{O}_1$ and $\mathcal{O}_2$ anticommute, $\{\mathcal{O}_1,\mathcal{O}_2\}=0$, we have
\begin{equation}
\mathcal{O}_1\mathcal{O}_2|\psi _{\pm }(\bm{k})\rangle =-\mathcal{O}_2\mathcal{O}_1|\psi _{\pm }(\bm{k})\rangle =\mp \iu \mathcal{O}_2|\psi _{\pm }(\bm{k})\rangle.
\end{equation}
Therefore, $\mathcal{O}_2\ket{\psi_\pm(\bm{k})}$ is also the eigenstate of $\mathcal{O}_1$ with eigenvalue $\mp \iu$ that is opposite to that of $\ket{\psi_\pm(\bm{k})}$. Namely, $\ket{\psi_\pm(\bm{k})}$ and $\mathcal{O}_2\ket{\psi_\pm(\bm{k})}$ are orthogonal states, implying  that the energy bands are  doubly degenerate. This combined symmetry can be broken by a net magnetization induced by an external magnetic field.

For the MX-II state that is obtained in the presence of a finite magnetic field along the $z$ axis, the combined symmetry of $\mathcal{O}_1$ and $\mathcal{O}_2$ is broken. Consequently, the double degeneracy is no longer present in the whole BZ. However, it is still present along the boundary of the folded first BZ, which is protected by the symmetries $\mathcal{O}_3$ and $\mathcal{O}_4$.

The double degeneracy along the BZ boundary $\bm{k}=(\pi/L,k_y)$ is protected by $\mathcal{O}_3$, where $L$ is the lattice constant of the magnetic superlattice. Because $\mathcal{O}_3^2=e^{- \iu k_x L}$, $\mathcal{O}_3$ becomes an antiunitary operator at $k_x=\pi/L$ where $\mathcal{O}_3^2=-1$.
Moreover, $\bm{k}=(\pi/L,k_y)$ is invariant under the action of $\mathcal{O}_3$ that changes $(k_x,k_y)\rightarrow (-k_x,k_y)$.
Therefore, there is a Kramers degeneracy of energy bands along the BZ boundary $\bm{k}=(\pi/L,k_y)$ [see Fig.~\ref{fig:symmetry}(b)]. This symmetry can be broken by a net magnetization along the $y$ direction.

For the same reason, the double degeneracy along the BZ boundary $\bm{k}=(k_x, \pi/L)$ is protected by $\mathcal{O}_4$ [see Fig.~\ref{fig:symmetry}(c)]. This symmetry can be broken by a net magnetization along $x$ direction.

We can further generalize the analysis to 3D systems, where we couple electrons to vertically stacked MX-II states. Similarly to the 2D case, there are four relevant symmetries at zero field:
\{$\mathcal{O}_1 \overline{\mathcal{M}}_z$,
$\mathcal{O}_2\overline{\mathcal{M}}_z$,
$\mathcal{O}_3\overline{\mathcal{M}}_z$,
$\mathcal{O}_4\overline{\mathcal{M}}_z$\},
where $\overline{\mathcal{M}}_z$ is the space reflection under which $z\rightarrow -z$.

These symmetries protect the band degeneracy at different parts of the 3D BZ:

(1) $\mathcal{O}_1 \overline{\mathcal{M}}_z$ and $\mathcal{O}_2\overline{\mathcal{M}}_z$ together protect the double degeneracy of all energy bands over the whole BZ. The combined symmetries can be broken by a net magnetization or magnetic field in any direction [Fig.~\ref{fig:symmetry_3d}(a)].

(2) $\mathcal{O}_3\overline{\mathcal{M}}_z$ protects the double degeneracy of energy bands over the BZ surface at $\bm{k}=(\pi/L,k_y,k_z)$. This symmetry can be broken by a net magnetization or magnetic field along $y$ direction [Fig.~\ref{fig:symmetry_3d}(b)].

(3) $\mathcal{O}_4\overline{\mathcal{M}}_z$ protects the double degeneracy of energy bands over the BZ surface at $\bm{k}=(k_x,\pi/L,k_z)$. This symmetry can be broken by a net magnetization or magnetic field along the $x$ direction [Fig.~\ref{fig:symmetry_3d}(c)].

Now we consider a transverse field in the $y$ direction that induces a net magnetization in the same direction and breaks the $\mathcal{O}_1 \overline{\mathcal{M}}_z$, $\mathcal{O}_2\overline{\mathcal{M}}_z$, and $\mathcal{O}_3\overline{\mathcal{M}}_z$ symmetries. According to the analysis above, the remaining $\mathcal{O}_4\overline{\mathcal{M}}_z$ symmetry protects the double degeneracy of the energy bands only at the surface of the BZ: $\bm{k}=(k_x,\pi/L,k_z)$ [Fig.~\ref{fig:symmetry_3d}(c)].
Further inclusion of the SOC in the interlayer hopping
\begin{align}
&\quad t_z \sum_i \sum_{\alpha\beta} \left( c_{i+\hat{z},\alpha}^\dagger \sigma_{\alpha\beta}^x c_{i,\beta} + \mathrm{H.c.} \right) \nonumber \\
&= \sum_{k_z} 2t_z \cos(k_z) \sum_{\alpha\beta}\sum_\gamma c_{k_z,\gamma,\alpha}^\dagger \sigma_{\alpha\beta}^x c_{k_z,\gamma,\beta},\label{tz}
\end{align}
breaks the $\mathcal{O}_4\overline{\mathcal{M}}_z$ symmetry ($\gamma$ is the index that labels different sublattices of the  magnet supercell). However, Eq.~\eqref{tz} vanishes at $k_z=\pm \pi/2$ resulting in two nodal lines at $\bm{k}=(k_x,\pi/L,\pm\pi/2)$ [Fig.~\ref{fig:Weyl}(a) in the main text].

To further reduce the nodal lines to Weyl points, we also include  SOC in the intralayer hopping,
\begin{equation}
t_x \sum_i \sum_{\alpha\beta} \left(c_{i+\hat{x},\alpha}^\dagger \sigma_{\alpha\beta}^x c_{i\beta} +\mathrm{H.c.} \right),
\end{equation}
which leads to the Weyl points shown in Fig.~\ref{fig:Weyl}(b) in the main text.

\bibliographystyle{apsrev4-1}
\bibliography{ref}

\end{document}